\documentclass[prb,showpacs,twocolumn,aps,superscriptaddress]{revtex4-1}

\usepackage{makeidx}
\usepackage{amsmath}
\usepackage{amssymb}
\usepackage{graphicx}
\usepackage{times}
\usepackage{graphics}
\usepackage{bm}
\usepackage{color}

\def\mbf#1{\mathbf{#1}}
\newcommand{\nn}{\nonumber}
\newcommand{\trl}{\tilde{R}_\ell}

\begin{document}

\title{Localization of spin waves in disordered quantum
rotors}
\author{Alexei Andreanov}
\affiliation{Max-Planck-Institut f\"{u}r Physik komplexer Systeme,
N\"{o}thnitzer Stra{\ss}e 38, D-01187 Dresden, Germany}
\author{Andrei A. Fedorenko}
\affiliation{CNRS UMR5672 - Laboratoire de Physique de l'Ecole Normale
Sup{\'e}rieure de Lyon, 46, All{\'e}e d'Italie, 69007 Lyon, France}

\date{\today}

\begin{abstract}
We study the dynamics of excitations in a system of $O(N)$ quantum rotors
in the presence of random fields and random anisotropies.
Below the lower critical dimension $d_{\mathrm{lc}}=4$  the system exhibits a
quasi-long-range order with a power-law decay
of correlations.  At zero temperature  the
spin waves are localized at the length scale $L_{\mathrm{loc}}$ beyond which
the quantum tunneling is exponentially suppressed
$ c \sim e^{-(L/L_{\mathrm{loc}})^{2(\theta+1)}}$.
At finite temperature $T$ the spin waves propagate by thermal activation
over energy barriers that scale as $L^{\theta}$.
Above $d_{\mathrm{lc}}$ the system undergoes an order-disorder phase transition
with activated dynamics such that the relaxation time grows with the correlation
length $\xi$ as $\tau \sim e^{C \xi^\theta/T}$ at finite temperature
and as $\tau \sim e^{C' \xi^{2(\theta+1)}/\hbar^2}$ in the vicinity of
the quantum critical point.

\end{abstract}

\maketitle

\section{Introduction}
Localization of excitations in disordered quantum systems has been attracting
considerable interest during the last several decades.
While single particle localization is rather well understood
within the standard theory of Anderson localization,~\cite{evers2008anderson}
localization of interacting particles is a much more
complicated  problem where many questions remain
open.~\cite{basko2006metal,aleiner2010finite,Yu2013localization}
Recently, excited many-body localized 
eigenstates were
studied in the random field Heisenberg spin-$\frac12$ chain
using exact diagonalization~\cite{pal2010many} and
in the random anisotropy \textit{XXZ} spin-$\frac12$ chain
by applying a dynamical real space
renormalization group.~\cite{vosk2013many}
It was found that
the many-body localized states in closed quantum systems
with quenched randomness
share many properties
with quantum glasses, \textit{e.g.}, they fail to thermally equilibrate
and  break ergodicity.
It was argued that such systems can be described by an infinite-randomness
fixed point (FP) with an infinite dynamic
critical exponent.~\cite{Yu2013localization,pal2010many,vosk2013many}
Unlike fermions, bosons can condense into a superfluid
state with long-range order so that interactions are intrinsically unavoidable.
The presence of disorder can
suppress the phase  coherence of the bosons and
localize them collectively in a compressible Bose glass with a gapless
energy spectrum~\cite{fisher1989boson,nelson1992boson,giamarchi1996variational,fedorenko2008elastic} or in an
incompressible Mott
glass.~\cite{yu2012bose,orignac1999possible,giamarchi2001competition}
The
zero-temperature
superfluid-insulator transition
in two-dimensional disordered hard-core bosons has been recently studied using a
spin-wave approach.~\cite{zuninga2013bose}
A mobility edge in the spin-wave excitation spectrum has been found
at a finite frequency that vanishes in the Bose glass phase.
The connection between
the Bose (Mott) glass and disordered elastic systems has been known for long
time.~\cite{nelson1992boson,giamarchi1996variational,%
fedorenko2008elastic}
Recently, a mapping of the leading order perturbation theory for boson
Green's functions to a directed polymer in random media has
been proposed for studying the insulating phase of
charged hard-core bosons.~\cite{gangopadhyay2013magnetoresistance}

In this paper we investigate the dynamics of a $d$-dimensional system of
$O(N)$ quantum rotors in  the presence of random fields and random anisotropies.
This model shares many properties with the aforementioned systems
but allows for an analytical study using the functional renormalization group (FRG)
that was originally developed for disordered elastic systems such as the directed
polymer in random
media.~\cite{fisher1986interface,nattermann1992dyanmics,chauve2001renormalization,fedorenko2006statics}
The FRG reveals that the behavior of the disordered quantum rotors
is controlled by a quasiclassical zero-temperature FP.
In the real space renormalization group treatment of
spin chains one fixes the temperature and
the Planck constant so that the renormalized disorder strength grows approaching
an infinite-randomness FP. In our FRG scheme we fix the disorder strength
near the FP but allow the temperature and the effective Planck constant to flow
to zero. Both parameters turn out to be
dangerously irrelevant like the temperature in the
random field Ising model.~\cite{fisher1986scaling} This  drastically
changes the dynamic scaling picture that one could expect from a naive
RG treatment.~\cite{dutta1998quantum,senthil1998properties}
The appearance of nonanalyticity in the FRG flow  prevents the system from
equilibration by inducing activated dynamics
with diverging  barriers at finite temperature and localization
at zero temperature.
This mechanism is to some extent similar to the one behind the classical
and quantum creep of disordered elastic systems at small driving
forces.~\cite{chauve2000creep,gorokhov2002creep}

The paper is organized as follows: We introduce the model in Sec.~\ref{sec:model}
and apply the FRG in Sec.~\ref{sec:frg}. In Sec.~\ref{sec:loc-ad} we discuss
the localization properties of excitations in the quasi-long-range order
(QLRO) phase below the lower critical dimension. Section~\ref{sec:transition}
is devoted to the activated  dynamics at the order-disorder transition above
the lower critical dimension. The Appendices present the technical details of
the derivation of the FRG flow equations.

\section{Model}\label{sec:model}
The Hamiltonian of interacting quantum rotors on a $d$-dimensional hyper-cubic
lattice with lattice constant~$b$ can be written as
\begin{equation}
	\label{eq-HR-1}
	\mathcal{H}_0=\frac{1}{2I} \sum_i \hat{\mathbf{L}}_i^2
- \sum_{\langle i,j \rangle} J_{ij} \hat{\mathbf{n}}_i \hat{\mathbf{n}}_j,
\ \ \ \ \ \hat{\mathbf{n}}_i^2=1,
\end{equation}
where the operator $\hat{\mathbf{n}}_i$ is a $N$-dimensional unit-length
vector  representing the orientation of the rotor
on site $i$.
$\hat{\mathbf{L}}_{i}$ is the angular momentum operator whose
$N(N-1)/2$ components are defined as
$\hat{L}_{i\mu\nu}= \hat{n}_{i\mu} \hat{p}_{i\nu} -\hat{n}_{i\nu} \hat{p}_{i\mu}$.
The momentum operator of each rotor with the moment of inertia $I$ satisfies
the commutation relations
$[\hat{n}_{i\mu},\hat{p}_{j\nu}]= i \hbar \delta_{ij} \delta_{\mu\nu}$.
The first term in~\eqref{eq-HR-1} is the kinetic energy of the rotor with
the moment of inertia $I$.
In the case of randomly distributed
exchange  interactions
$J_{ij}$ the system forms a strong quantum glass which has been studied
mainly in the limit of infinite range interactions using
$1/N$-expansion.~\cite{ye1993solvable} The limit of $N=1$ is expected to be in the
same universality class as the Ising model in a transverse field
whose glass phase is critical everywhere and exhibits gapless collective
excitations in the long-range interaction limit.~\cite{andreanov2012longrange}
Here we assume that all $J_{ij}=J$ and restrict the sum ${\langle i,j \rangle}$
to nearest neighbors. Instead of the random exchange interactions
we introduce random fields and random anisotropies as
 $\mathcal{H}=\mathcal{H}_0+\mathcal{H}_{\mathrm{RF}}+\mathcal{H}_{\mathrm{RA}}$,
where
$ \mathcal{H}_{\mathrm{RF}}=
-\sum_{i} \mathbf{h}_i \cdot \hat{\mathbf{n}}_i $ and
$ \mathcal{H}_{\mathrm{RA}}=
-\sum_{i}(\mathbf{d}_i \cdot \hat{\mathbf{n}}_i )^2$ with
randomly oriented vectors $\mathbf{h}_i$ and $\mathbf{d}_i$.
In the continuum limit this model can be rewritten as
an $O(N)$ quantum-mechanical nonlinear
$\sigma$-model (QNL$\sigma$M) with the partition function
$\mathcal{Z}=\int \mathcal{D} \mathbf{n}\, %
 \delta (|\mathbf{n}|-1) e^{-\mathcal{S[\mathbf{n}]}/\hbar}$
and the imaginary time action
\begin{eqnarray} \label{eq-action-continuum-l}
&& \!\!\!\!\!\!	\mathcal{S}\left[ \mathbf{n} \right] =
   \frac{\rho_0 }2  \int_{\tau,x} \left[ \frac{1}{c_0^2}
   \left(\partial_\tau \mathbf{n}(\tau,x)\right)^2
   + \left(\nabla \mathbf{n}(\tau,x)\right)^2\right] \nn \\
&& \ \ \ \ \	- \int_{\tau,x} \sum\limits_{\mu=1}^{\infty}\sum\limits_{i_1\cdots i_{\mu}}
h^{(\mu)}_{i_1\cdots i_ {\mu}}(x) n_{i_1}(\tau,x)\cdots n_{i_{\mu}}(\tau,x), \ \ \
\end{eqnarray}
where we have introduced the shorthand notations
$\int_\tau:=\int_0^{\hbar/T} d \tau$ and $\int_{x}:=\int d^d x$. Here
$\rho_0=b^{2-d}J$ is the bare stiffness constant,
$c_0= b \sqrt{J/I}$  the bare spin-wave
velocity and $T$  the temperature. The UV cutoff $\Lambda_0=2\pi/b$ is
imposed in~\eqref{eq-action-continuum-l}.
The $O(N)$ QNL$\sigma$M arises as an effective theory for the
low energy degrees of freedom in several correlated quantum systems.
For instance, the $O(2)$ model  describes Cooper
pairs of electrons in a superconducting Josephson junctions  array
and ultra-cold atoms in an optical lattice.~\cite{cazalilla2011one}
The $O(3)$ model describes a quantum spin-$S$ antiferromagnet
in the large-$S$ limit.~\cite{chakravarty1989two}
The $O(5)$ QNL$\sigma$M was suggested for the
unified  low-energy  theory of the antiferromagnetic and superconducting
phases in the high-$T_c$ superconductors.~\cite{demler2004so5}
The renormalization of the original model~\eqref{eq-HR-1}
with random fields $h^{(1)}$ and/or
anisotropies  $h^{(2)}$
generates
the higher rank anisotropies  $h^{(\mu)}$,
which we incorporated in the second line
of~\eqref{eq-action-continuum-l} from
the beginning.~\cite{fisher1985random,fedorenko2007long-range}
%
The RG flow preserves the symmetry
with respect to inversion $\mathbf{n}\to - \mathbf{n}$,
so we will use the notation of random
anisotropy (RA) for the systems respecting this symmetry  and
random field (RF) for the rest.
The bare $\mu$th rank anisotropies can be taken to be
Gaussian distributed  with zero mean and cumulants
\begin{equation}
	\overline{h^{(\mu)}_{i_1\cdots i_{\mu}}(x) h^{(\nu)}_{j_1\cdots j_{\nu}}(x')}
 =  \delta^{\mu\nu}\delta_{i_1 j_1} \cdots \delta_{i_{\mu} j_{\nu}}
r^{(\mu)}\delta(x-x').
\end{equation}
We use the replica trick to average over disorder.
Introducing $n$ replicas of the original system we obtain
the replicated action
\begin{eqnarray}
  \mathcal{S}_n\left[ \{\mathbf{n} \} \right] &=& \frac{\rho_0 }2 \sum_{a=1}^n
   \int_{\tau,x} \left[ \frac{1}{c_0^2}
   \left(\partial_\tau \mathbf{n}_a(\tau,x)\right)^2
   + \left(\nabla \mathbf{n}_a(\tau,x)\right)^2\right] \notag\\
	&& -  \frac{1}{2\hbar} \sum_{a,b=1}^n  \int_{\tau,\tau', x} \,
 R\big(\mathbf{n}_a(\tau,x)\cdot\mathbf{n}_b(\tau',x)\big),\ \ \ \ \
	\label{eq-action-rep-2}
\end{eqnarray}
where we have introduced $R(z)=\sum_{\mu}r^{(\mu)} z^{\mu}$, which is
defined for $-1\le z \le 1$.
This function is even for the RA model
and has no symmetry for the RF model.
The properties of the original disordered
system~\eqref{eq-action-continuum-l} can be extracted in
the limit $n\to 0$.
The Imry-Ma arguments suggest that true long-range order
is absent in our model for $d<4$, \textit{i.e.} $d_{\mathrm{lc}}=4$
is the lower critical
dimension. However, a quantum QLRO
can survive at low enough temperature, similarly to the QLRO in the
classical Heisenberg model.~\cite{feldman2000quasi}
In the QLRO phase the local order slowly changes in
space, leading to a power-law decay of correlations that justifies
the description of the dynamics in terms of spin-wave
excitations.

\section{Functional renormalization group} \label{sec:frg}
To get access to the low-$T$ phase we renormalize the
action~\eqref{eq-action-rep-2} using a momentum-shell method in which
iterative integrations over fast modes with wavevectors between the
bare cutoff  $\Lambda_0$ and the running cutoff $\Lambda_\ell=\Lambda_0 e^{-\ell}$
generate the RG flow equations.
Dimensional analysis of the action~(\ref{eq-action-rep-2}) shows that
all the derivatives of $R(z)$ at $z=0$ are relevant operators. Thus, one needs
to follow the renormalization of the entire function $R(z)$.
It is convenient to express the flow equations in terms of the reduced
running quantities:
\begin{eqnarray}
\trl(\phi)&=& K_d R_\ell(z)\rho_{\ell}^{-2} \Lambda_\ell^{d-4}, \label{eq-red-R}  \\
\tilde{\hbar}_\ell&=& K_d \hbar \rho_{\ell}^{-1} \Lambda_\ell^{d-1}, \label{eq-red-hbar} \\
\tilde{T}_\ell&=& K_d T_\ell \rho_{\ell}^{-1} \Lambda_\ell^{d-2}, \label{eq-red-T}
\end{eqnarray}
%
where $z=\cos\phi$ and $K_d$ is the surface of  the unit sphere in $d$- dimensions divided by
$(2\pi)^d$. The function $\tilde{R}_\ell(\phi)$ is $\pi$-periodic for
the RA and $2\pi$-periodic for the RF model.
We expand the action around a locally ordered state and neglect the
possible presence of
topological defects that can modify the behavior of the
system.~\cite{proctor2014random}
We split the local order parameter
$\mbf{n}_a=({\sigma}_a,\bm{\pi}_a)$
into the component ${\sigma}_a=\sqrt{1-{\bm{\pi}}_a^2}$ aligned along the
locally preferred direction and the $(N-1)$-component vector $\bm{\pi}_a$
perpendicular to it.
We decompose the latter into slowly and rapidly varying parts $\bm{\pi}_{a}^{<}$
and  $\bm{\pi}_{a}^{>}$ with
the momentum modes $0<q<\Lambda_\ell$ and $\Lambda_\ell<q<\Lambda_0$, respectively.
Integrating out the fast
fields $\bm{\pi}_{a}^{>}$
and allowing for the rescaling of the slow fields
$\bm{\pi}_{aR} (x) =\zeta \bm{\pi}_{a}^{<} (x)$ with
\begin{equation}\label{eq-zeta}
\zeta= 1 + \frac12 (N-1) \trl''(0) \ell + O(\trl^2),
\end{equation}
we obtain the one-loop flow equations for the effective temperature $\tilde{T}_{\ell}$
and the Planck constant $\tilde{\hbar}_{\ell}$ (see Appendix~\ref{sec:A}
for more details)
\begin{equation}
 \partial_{\ell} \ln \tilde{T}_{\ell} = 1+\partial_{\ell} \ln \tilde{\hbar}_{\ell}
    = 2-d - (N-2) \tilde{R}''_{\ell}(0), \label{eq-flow-T-h}
\end{equation}
and for the disorder correlator
\begin{eqnarray}
&& \partial_{\ell} \trl(\phi)= \varepsilon \trl(\phi)
+ \trl''(\phi)[\Gamma_{\ell} - \trl''(0)]
+ \frac12 [\trl''(\phi)]^2
  \nn \\
&& + (N-2)\left( \frac{\trl'(\phi)^2}{2\sin^2 \phi} \right.
  + \left.
\left[\frac{\trl'(\phi)}{\tan \phi}+
 2\trl(\phi)\right][\Gamma_{\ell} - \trl''(0)] \right). \nn \\
  \label{eq-FRG-1-n}
\end{eqnarray}
Here we introduced  $\varepsilon=4-d$ and the boundary layer width
\begin{equation}\label{eq-Gamma-1}
\Gamma_{\ell}=\frac12 c_{\ell} \tilde{\hbar}_{\ell} \coth \left[
\frac{c_\ell \tilde{\hbar}_{\ell} }{2 \tilde{T}_{\ell}}\right]
= \left\{\begin{array}{cl}
\displaystyle \tilde{T}_{\ell} \quad& \mbox{if}~ \tilde{\hbar}_{\ell} \to 0 ,\\
\displaystyle \frac12 c_\ell \tilde{\hbar}_{\ell}  \quad& \mbox{if}~ \tilde{T}_{\ell} \to 0, \\
\end{array} \right.
\end{equation}
that describes the joint effect of thermal and quantum fluctuations on the disorder
correlator flow.
Disorder breaks the Lorentz invariance of the clean system and
renormalizes  the spin-wave velocity
\begin{equation}
\partial_{\ell} \ln c_{\ell} = - \frac16\left[(N+1)\trl^{(4)}(0)
 + (N-2)\trl''(0)\right] \label{eq-flow-c}
\end{equation}
similarly to the stiffness constant in disordered
elastic systems with broken statistical tilt
symmetry.~\cite{fedorenko2008elastic,giamarchi1996variational}

Assuming that the running disorder correlator reaches an attractive FP of the
flow equation~\eqref{eq-FRG-1-n}
one might naively  conclude  from~\eqref{eq-flow-T-h} and \eqref{eq-flow-c} that
the system exhibits a usual critical scaling behavior.
However, the more accurate analysis
presented below for $d<d_{\mathrm{lc}}=4$ and $d>d_{\mathrm{lc}}$
shows that this is not the case.

%
\begin{figure}
\includegraphics[width=7cm]{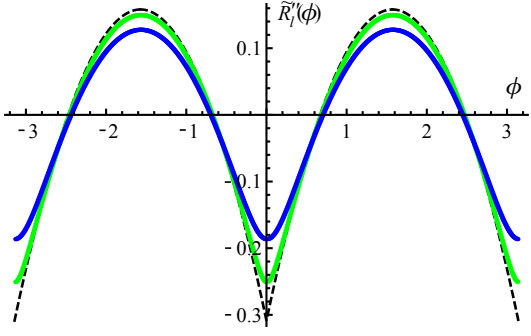}
\caption{(Color online) The $\pi$ - periodic FP solution $\tilde{R}''(\phi)$
describing the QLRO phase in the 3D $O(3)$ RA model.
The dashed line is the zero-$\Gamma$ FP, the solid green and blue lines
with a rounded cusp are the finite $\Gamma$ FPs for $\Gamma=0.05$ and $\Gamma=0.1$.
 }
  \label{fig:rafpn3-1}
\end{figure}
%

\section{Localization and activated dynamics} \label{sec:loc-ad}

We start the analysis of Eqs.~(\ref{eq-zeta})-(\ref{eq-flow-c})
for $d<d_{\mathrm{lc}}=4$ by studying the flow of the disorder correlator for
$\Gamma_{\ell}=0$, \textit{i.e.} neglecting thermal and quantum fluctuations.
For concreteness we take a smooth $\pi$- periodic bare
correlator $\tilde{R}_0(\phi)=\gamma \cos^2 \phi$ (the RA universality
class).
The flow equations for the first derivatives of $\trl(\phi)$
at $\phi=0$, which follow from~\eqref{eq-FRG-1-n}, imply that the
renormalized $\tilde{R}^{(4)}_{\ell}(0)$ diverges at the  finite scale
\begin{equation}
\ell_c\approx \frac1{\varepsilon} \ln\{1+3\varepsilon/[8\gamma(N+7)]\}.
\end{equation}
Beyond this scale the running disorder correlator
becomes non-analytic at $\phi=0$:
the second derivative develops a cusp,
$\tilde{R}'''_{\ell}(0^+)\neq 0$ for $\ell>\ell_c$.
Then the renormalized
disorder correlator $\trl(\phi)$ rapidly approaches a non analytic
FP solution $\tilde{R}^*(\phi)$ with
$\tilde{R}^{*\prime\prime\prime}(0^+)\neq 0$ and finite $\tilde{R}^{*(4)}(0^+)$.
The stable non-analytic FP solution exists for $2\le N \le N_c $ with
$N_c=2.835$ for RF and $N_c=9.441$ for RA.
These values are close to their classical
limits.~\cite{tissier2006two,ledoussal2006random,fedorenko2012random}
For instance,  the $O(2)$ model has the FPs with
$\tilde{R}^{*\prime\prime}(0)=-\phi_0^2\varepsilon/36$, where $\phi_0=\pi$ and
$\phi_0=2\pi$ for the RA and RF models, respectively. The $O(3)$ and $O(4)$ RA models
have the FPs with $\tilde{R}^{*\prime\prime}(0)\approx -0.309\varepsilon$
and $\tilde{R}^{*\prime\prime}(0)\approx -0.358\varepsilon$.
The numerical $\Gamma=0$ RA FP solution for $N=3$ is shown
in Fig.~\ref{fig:rafpn3-1}.

The numerical analysis of the full FRG flow~(\ref{eq-flow-T-h})-(\ref{eq-flow-c})
shows
that the running disorder correlator $\tilde{R}_\ell$ can be replaced for $ß\ell> \ell_c$ by
the FP point solution of the flow equation~\eqref{eq-FRG-1-n}
at fixed $\Gamma_{\ell}$.
For a finite but small $\Gamma_{\ell}$ this FP solution uniformly
approaches the zero-$\Gamma$
FP solution everywhere except for the extreme points~(see Fig.~\ref{fig:rafpn3-1}).
The physically most relevant region is the boundary layer around
$\phi=0$ which has the width of order $\Gamma_{\ell}$.
Within the boundary layer the cusp
of the zero-$\Gamma$ FP solution $\tilde{R}^{*\prime\prime}(\phi)$
is rounded by thermal and quantum fluctuations. Indeed,
since $\Gamma_{\ell}$ flows towards zero the second derivative
$\tilde{R}''_{\ell}(0)$ approaches $\tilde{R}^{*\prime\prime}(0)\neq 0$ while
$\tilde{R}^{(4)}_{\ell}(0)$ diverges, and thus, remains different from
$\tilde{R}^{*(4)}(0^+)$ for arbitrary  small but finite $\Gamma_{\ell}$.
This results in activated dynamic scaling
similar to that found in the random
transverse field Ising model~\cite{young1996numerical,monthus2012random}
and may lead to different behavior of averaged
and typical correlations and multifractality.~\cite{fedorenko2014gaussian}
In particular the averaged connected and disconnected correlations scale
differently:
\begin{eqnarray}
G_{\textrm{con}}(x) \sim 1/x^{d-2+\eta}, \ \ \ \
G_{\textrm{dis}}(x) \sim 1/x^{d-4+\bar{\eta}} \label{eq-GGG}
\end{eqnarray}
with the exponents
\begin{eqnarray}
\eta&=&-\tilde{R}^{*\prime\prime}(0), \\
\bar{\eta}&=&\varepsilon-(N-1)\tilde{R}^{*\prime\prime}(0),
\end{eqnarray}
which can be extracted from the rescaling factor (\ref{eq-zeta}) at the FP
(for details see Ref.~\onlinecite{fedorenko2012random}).
The algebraic decay of correlators implies that the spectrum of excitations remains
gapless in the whole quantum QLRO phase.  This is in contrast to the
pure model in the disordered phase with a gap in the energy spectrum
that vanishes only at the transition to the ordered state:
the quantum transition occurs when the bare effective coupling constant
$g_0 =c_0 \tilde{\hbar}_0$ crosses
a nontrivial FP $g^*=2(d-1)/(N-2)$ at zero temperature while the thermal transition
takes place along the separatrix controlled by a thermal FP $g^*=0$ and
$\tilde{T}^*=(d-2)/(N-2)$.~\cite{chakravarty1989two}

\begin{figure}
\includegraphics[width=7cm]{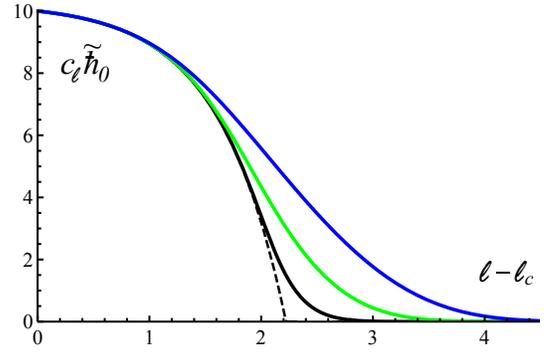}
\caption{(Color online) The renormalized spin-wave velocity in the 3D $O(3)$ RA model
as a function of $\ell-\ell_c$ for different $T$ and the initial condition
for the bare coupling constant $g_0=c_0\tilde{\hbar}_0=10$.
The dashed line corresponds to $\tilde{T}_0=0$; the solid black, green, and blue
lines to $\tilde{T}_0=0.2;0.5;1$.
 }
  \label{fig:spin-wave}
\end{figure}

To find the flow of the disorder correlator in the boundary layer we
expand the flow equation~\eqref{eq-FRG-1-n} in small $\phi$ for fixed $\Gamma_\ell$.
To lowest order in $\Gamma_\ell$ this gives
$\trl''(0)\approx\tilde{R}^{*\prime\prime}(0)$ and
\begin{eqnarray}
\trl^{(4)}(0)\approx 6\Omega/[\Gamma_{\ell} (N+1)]
\end{eqnarray}
with the universal constant
\begin{equation}
\Omega=  \frac12\tilde{R}^{*\prime\prime}(0)
 [\tilde{R}^{*\prime\prime}(0)(N-2)-\varepsilon].
\end{equation}
The flow for $\ell<\ell_c$ is analytic and leads to renormalization
of the bare  parameters $\tilde{T}$, $\tilde{\hbar} $ and $c$ by factors of order~$1$.
Neglecting  the latter
we obtain from~\eqref{eq-flow-T-h} that $\tilde{T}_{\ell} = \tilde{T}_0 e^{-\theta (\ell-\ell_c)}$
and $\tilde{\hbar}_{\ell} = \tilde{\hbar}_0 e^{-\theta_\hbar (\ell-\ell_c)}$.
The exponents $\theta$ and $\theta_\hbar$ are given by
\begin{equation}
\theta=\theta_\hbar-1=d-2+ (N-2) \tilde{R}^{*\prime\prime}(0),
\end{equation}
to one loop order.
Note that the exponent $\theta_\hbar$ coincides with the exponent $\theta$ in the
corresponding classical system in $d+1$ dimensions with columnar disorder.
We conjecture that the relation $\theta_\hbar=1+\theta$ holds to all orders.
Substituting the disorder correlator derivatives into the boundary
layer to the spin-wave velocity flow~\eqref{eq-flow-c} and omitting
the subdominant terms we find
\begin{eqnarray}
\partial_{\ell} \ln c_{\ell} = - \frac{\Omega}{\Gamma_l}. \label{eq-flow-c-2}
\end{eqnarray}
In the classical limit
$\tilde{\hbar} \to 0$, $\tilde{T} \to \infty$ the rounding of the
cusp in the boundary
layer is governed by thermal fluctuations, $\Gamma_\ell \approx \tilde{T}_\ell$.
Neglecting renormalization of the spin-wave velocity $c$ below the scale  $\ell_c$
we arrive at
\begin{eqnarray} \label{eq-cell}
c_\ell = c_0 e^{-\frac{\Omega}{\tilde{T}_0 \theta}[ e^{\theta (\ell-\ell_c)}-1]}.
\end{eqnarray}
Thus, in the classical regime the low frequency spin-waves propagate
via thermal activation over energy barriers that grow with the length scale
$L=\Lambda_0^{-1}e^{\ell}$ as $L^{\theta}$. We believe that this result is
also applicable to the classical $O(N)$ models with Langevin dynamics where
$c$ has to be replaced by the kinetic coefficient.~\cite{fedorenko2006critical}
While early numerical works~\cite{itakura2003frozen} confirmed
a power-law decay of correlations in the classical $O(N)$ models, recent numerical
simulations~\cite{proctor2014random} suggested that the presence
of topological defects can lead to an exponential decay of correlations on
scales larger than the average distance between the defects. Thus, there
is a possibility for a scenario when the dynamics is described by~\eqref{eq-cell}
while the algebraic decay of correlations is screened by
the topological defects whose relaxation time is very large.
In the opposite limit of  $T \to 0$,
the spin-wave velocity
vanishes at a finite length scale
$L_{\mathrm{loc}}=\Lambda_0^{-1}e^{\ell_{\mathrm{loc}}}$ with
\begin{eqnarray}
\ell_{\mathrm{loc}}-\ell_c = \frac1{\theta_\hbar}
\ln \left[ 1+  \frac{c_0 \tilde{\hbar}_0 \theta_\hbar }{2\Omega}
 \right]. \label{eq-loc-length}
\end{eqnarray}
This means that the magnon excitations cannot propagate on distances larger
than this scale which can be interpreted as the zero temperature
spin-wave localization length.
The spin and energy transport is strongly suppressed beyond this length scale
leading to failure of quantum thermalization.
The renormalized  spin-wave velocity
computed from numerical integration
of the flow equation~\eqref{eq-flow-c-2} for different temperatures
is shown in Fig.~\ref{fig:spin-wave}.
For finite but small temperature $T$ one
can define an effective localization length
\begin{eqnarray}
L_T \approx  L_{\mathrm{loc}}
\left[1+ \frac{\tilde{T}_0 \theta}{\Omega}
\ln \left[ c_0 \Lambda_0 \tau_{\mathrm{exp}}\right] \right]^{1/\theta},
\end{eqnarray}
beyond  which the activated dynamics can be neglected on the time scale
of experiment $\tau_{\mathrm{exp}}$.

In deriving~\eqref{eq-Gamma-1} we assumed that the $\Gamma_\ell$
is determined exclusively by the low-frequency part of the spectrum.
We now show that taking into account the renormalization of the high-frequency
part of the excitation spectrum
leads to an extremely small but finite spin-wave velocity in the low frequency
limit even at zero temperature. To see that we generalize the bare part of
the effective action~\eqref{eq-action-rep-2} at $T=0$ to
\begin{eqnarray}
  \mathcal{S}_n^{(0)} &=& \frac{\rho_0 }2 \sum_{a=1}^n
   \int_{-\infty}^{\infty} \frac{d \omega}{2\pi} \int\frac{d^d q}{(2\pi)^d}
    \left[ D(\omega) + q^2 \right]
|\mathbf{n}_a(\omega,q)|^2. \nonumber \\
	\label{eq-action-rep-3}
\end{eqnarray}
Such generalization does not modify the flow equations for the effective
Planck constant~\eqref{eq-flow-T-h} and disorder correlator~\eqref{eq-FRG-1-n}.
The zero-temperature boundary layer width is, however, now given by
\begin{eqnarray} \label{eq-Gamma-3}
\Gamma_\ell = \frac{\tilde{\hbar}_\ell}{\Lambda_\ell}
\int^{\infty}_{-\infty}\frac{d \omega}{2\pi}
\frac1{1+\tilde{D}_\ell(\omega)},
\end{eqnarray}
where we have defined
$ \tilde{D}_{\ell}(\omega) = \Lambda_\ell^{-2} D_{\ell}(\omega)$.
The flow of the spectrum $\tilde{D}(\omega)$
starting from an arbitrary phononlike spectrum $\tilde{D}_0(\omega)$
to one loop order reads (see Appendix~\ref{sec:B})
\begin{eqnarray} \label{eq-D-1}
\partial_\ell \tilde{D}_{\ell}(\omega) = 2\tilde{D}_{\ell}(\omega)
+\frac{2\Omega}{ \Gamma_{\ell}}
 \frac{\tilde{D}_{\ell}(\omega)}{1+\tilde{D}_{\ell}(\omega)}.
\end{eqnarray}
Here we have retained only the terms that are relevant in the
limit $\Gamma_\ell \to 0$.
The renormalized spectrum and the boundary layer width are solutions
of the self-consistent equations~\eqref{eq-Gamma-3} and \eqref{eq-D-1}.
Renormalization of the high frequency part of the spectrum~\eqref{eq-D-1}
contributes to the boundary layer width~\eqref{eq-Gamma-3} and
leads to an exponentially small spin-wave velocity
on scales $L>L_{\mathrm{loc}}$.
To see this we solve the spectrum flow equation~\eqref{eq-D-1} in the
high frequency region $\tilde{D}(\omega) \gg 1$:
\begin{eqnarray}
\tilde{D}_\ell(\omega) \approx e^{2\ell} \tilde{D}_0(\omega) + 2\Omega
\int_{\ell_c}^{l}\frac{d\ell'}{\Gamma_{\ell'}} e^{2(\ell-\ell')}.
\end{eqnarray}
Plugging this in~\eqref{eq-Gamma-3} and taking the bare spectrum as
$\tilde{D}_{0}(\omega)= \omega^2/(c_0^2 \Lambda_0^2)$ we integrate out
the high frequencies and obtain a Volterra-type integral equation for $\Gamma_\ell$.
For large $\ell$ it can be transformed into a differential equation:
\begin{eqnarray}
\frac{d}{d\ell} \left[\frac{c_0 \tilde{\hbar}_\ell
e^{-(\ell-\ell_c)}}{2\Gamma_\ell}\right]^2 =
\frac{2\Omega }{\Gamma_\ell}e^{-2(\ell-\ell_c)},
\end{eqnarray}
whose solution is
$ \Gamma_\ell = c_0^2\tilde{\hbar}_\ell ^2 \theta  /(4\Omega)$.
Using the flow equation~\eqref{eq-flow-c-2} we find the spin-wave velocity
contribution due to renormalization of the high frequency part of the spectrum
\begin{eqnarray} \label{eq-c-qq}
c(L) \sim \exp\left[-\frac{1+\theta}{2\theta} \left(
\frac{L}{L_{\mathrm{loc}}}\right)^{2(\theta+1)} \right],
\end{eqnarray}
which we have expressed in terms of the localization
length~\eqref{eq-loc-length} and using the relation
$\theta_\hbar=1+\theta$. Equation~(\ref{eq-c-qq}) shows that
the effect of the residual quantum tunneling is much weaker than
the effect of the thermal activation~(\ref{eq-cell}).

\section{Order-disorder transition} \label{sec:transition}
Above the lower critical dimension $d_{\mathrm{lc}}=4$ the quantum
model~\eqref{eq-action-continuum-l} undergoes an order-disorder transition
similar to that of the classical model.~\cite{feldman2002critical}
For $\varepsilon<0$ and $N>N_c$ the FRG equation~\eqref{eq-FRG-1-n}
has a FP solution which is unstable in a single direction, and thus,
describes the transition.
For instance, the $O(3)$ and $O(4)$ RF models have
the FPs with
$\tilde{R}^{*\prime\prime}(0)=-5.54|\varepsilon|$ and
$\tilde{R}^{*\prime\prime}(0)=-0.787|\varepsilon|$, respectively.
For $N>18$ the non analyticity of the RF FP  becomes weaker
than a linear cusp in $\tilde{R}''(\phi)$  and its value sticks to
$\tilde{R}^{*\prime\prime}(0) \approx -|\varepsilon|/(N-2)$.
The large $N$ behavior of the RA FP is given by
$\tilde{R}^{*\prime\prime}(0) \approx -|\varepsilon|(3N+40)/2(N-2)^2$.

The critical temperature $T_c(\Delta)$ is a function of the bare disorder
strength (e.g. $\Delta=r^{(1)}$ for RF and  $\Delta=r^{(2)}$ for RA) and
vanishes at the quantum critical point $T_c(\Delta^*)=0$.
The only positive eigenvalue $\lambda=|\varepsilon|$ does not depend on $N$
to one loop order and gives the critical exponent $\nu=1/\lambda$,
that describes the divergence of the  correlation length in the classical
regime $\xi\sim |T-T_c|^{-\nu}$ and at the quantum critical point
$\xi\sim |\Delta-\Delta^*|^{-\nu}$. The hyperscaling relation between $\nu$ and the
heat capacity exponent $\alpha$ is modified by the exponent $\theta$ as
$\nu(d-\theta)=2-\alpha$.
The averaged connected and disconnected correlation functions exhibit
the power-law behavior~(\ref{eq-GGG}) at the transition with the exponents
$\eta$ and $\bar{\eta}$
related by $\bar{\eta}=2+\eta-\theta$.
The critical dynamics
can be studied along the same lines as for the dynamics in the QLRO phase.
It turns out to be activated as well, with the typical relaxation time
\begin{equation}
\tau \sim e^{C \xi^\theta/T}
\end{equation}
in the classical regime
with $C=\Omega \rho_0 \Lambda_0^{\theta-d+2}/K_d\theta$
and the typical relaxation time
\begin{equation}
\tau \sim e^{C' \xi^{\Psi}/\hbar^2}
\end{equation}
with
$C'= 2 C^2 \theta/(1+\theta) $ and
\begin{eqnarray}
\Psi=2\theta_\hbar=2(\theta+1) \label{eq-Psi}
\end{eqnarray}
at the quantum critical point.
We expect the scaling relation~(\ref{eq-Psi})
to hold also in the Ising case.~\cite{senthil1998properties,anfuso2009random}
Note that the capability of the $\varepsilon$ expansion (or other perturbative
approaches such as a $2+\varepsilon$ expansion) to provide evidence for
the activated classical or quantum dynamics which is expected in the random
field spin systems from numerics and phenomenology has been much debated
in the literature.\cite{senthil1998properties}
Our results prove the power of the FRG method, in particular
its ability to capture the activated dynamics.

\section{Conclusion}
We have studied the dynamics of disordered interacting quantum rotors. We found
that the system
is controlled by a quasiclassical zero-temperature
(\textit{i.e.}, infinite randomness) FP with an infinite dynamic critical exponent.
Below the lower critical dimension $d_{\mathrm{lc}}=4$
the system has a quantum QLRO phase with a power-law
decay of correlations. At zero temperature  the spin-wave excitations are localized
on the length scale $L_{\mathrm{loc}}$ that prevents quantum thermalization.
For $T>0$ the spin-waves propagate via thermal activation
over the energy barriers 
which diverge in the thermodynamic dynamic limit so that the system never
thermally equilibrates.
These results, obtained for the 3D $O(2)$ RF and $O(3)$ RA models, can be relevant
for the
quantum dynamics  of the Bose glass and disordered quantum antiferromagnets.

Above the lower critical dimension the system of quantum rotors undergoes an
order-disorder phase transition with activated dynamics which is strongly
suppressed in the vicinity of the quantum critical point.

\begin{acknowledgments}
 We would like to thank
L.~Cugliandolo, J.~Wehr, M. Gingras, T.~Roscilde, D.~Carpentier, E.~Orignac,
P.~Le~Doussal, and K.~J.~Wiese for stimulating discussions. AAF acknowledges
support by ANR grants 13-JS04-0005-01 (ArtiQ) and 2010-BLANC-041902 (IsoTop).
\end{acknowledgments}

\appendix

\section{Derivation of the flow equations} \label{sec:A}

In order to derive the flow equations  we introduce the IR cutoff by imposing
a homogeneous external field $\mbf{h}$
which is linearly coupled to $\mbf{n}$.
In the limit of small temperature and weak disorder the system is fluctuating
around  the completely ordered state
in which all replicas of all spins align along the direction of $\mbf{h}$.
We split the order parameter
$\mbf{n}_a=({\sigma}_a,\bm{\pi}_a)$
into the $(N-1)$-component vector $\bm{\pi}_a$ which is perpendicular
to $\mbf{h}$ and the component ${\sigma}_a=\sqrt{1-{\bm{\pi}}_a^2}$
parallel to it.
Then the partition function can be rewritten as
\begin{eqnarray}
Z=\int \prod\limits_{a=1}^{n} {\cal D} \bm{\pi}_a
 \prod\limits_{\tau,x}^{}
\frac1{\sqrt{1-\bm{\pi}_a^2(\tau,x)}} e^{-\mathcal{S}_n[\bm{\pi}]/\hbar}
\end{eqnarray}
with the replicated action
\begin{eqnarray}
&& \!\!\! \mathcal{S}_n[\bm{\pi}]=
   \frac{\rho_0}{2} \int_{x,\tau} \sum_{a=1}^n \left\{ \frac1{c_0^2} \left[
   \left(\partial_{\tau} \bm{\pi}_a \right)^2
   +\frac{(\bm{\pi}_a \cdot \partial_{\tau} \bm{\pi}_a)^2}{(1-\bm{\pi}_a^2)}
      \right] \right.
     \nonumber \\
&& \ \ \ \ \ \ + \left.
   \left(\nabla \bm{\pi}_a \right)^2
   +\frac{(\bm{\pi}_a \cdot \nabla \bm{\pi}_a)^2}{(1-\bm{\pi}_a^2)}
     - h\, \sqrt{1-{\bm{\pi}}_a^2} \right\}   \nonumber \\
&&  \ \ \ \ \ \ -  \frac{1}{2\hbar} \sum_{a,b=1}^n   \int_{x,\tau,\tau'} \,
 {R}\big( \bm{\pi}_a(\tau,x)\cdot \bm{\pi}_b(\tau',x) \nn \\
&& \ \ \ \ \ \
   + \sigma_a(\tau,x) \sigma_b(\tau',x)
  \big).
     \label{eq:S-1}
\end{eqnarray}
We use the momentum shell method
developed in Refs.~\onlinecite{chakravarty1989two,nelson1977momentum} and
consider the
loop expansion in small $\hbar$ and $R$. To that end we express $\bm{\pi}_a$ as
\begin{eqnarray}
\bm{\pi}_a(\tau,x)=\sum\limits_{m=-\infty}^{\infty} \int\frac{d^d q}{(2\pi)^d}
 \bm{\pi}_a(\omega_m, q) e^{i \omega_m \tau + i q\cdot x},
\end{eqnarray}
where we have introduced the Matsubara frequencies $\omega_m=2\pi T m/\hbar$
with $m \in \mathbb{Z}$. We now decompose the fields $\bm{\pi}_a$
into slowly and rapidly varying parts as follows
\begin{equation}
\bm{\pi}_a(\omega_m, q) =
\left\{
\begin{array}{l}
  \bm{\pi}_{a}^{<}(\omega_m, q),  \ \ \  0<q<\Lambda_\ell. \\
  \bm{\pi}_{a}^{>}(\omega_m, q), \ \ \ \Lambda_\ell<q<\Lambda_0.
\end{array}
\right.
\end{equation}
Integrating out $\bm{\pi}_{a}^{>}$ and rescaling momenta by $e^{\ell}$
and the fields $\bm{\pi}_{a}^{<}$ by $\zeta$ we obtain the effective
action of the same form (\ref{eq:S-1})  which involves only
$\bm{\pi}_{a}^{<}$ and the new parameters $T'$, $c'$, $\hbar'$,
$h'$ and $[R(z)]'$.
It is convenient to introduce $\hat{R}(\phi)={R}(z)$   with $z=\cos\phi$.
The bare disorder correlator ${R}(z)$ is an analytic function of $z$ for
$-1\le z \le 1 $. However, the renormalized disorder correlator
becomes nonanalytic around $z=1$ and as can be checked \textit{a posteriori} by
solving the flow equation it has the following expansion:\cite{tissier2006two}
\begin{eqnarray}
{R}(z)&=&{R}(1)+{R}'(1)(z-1)+\frac{{a}_1}3 [2(1-z)]^{3/2}\nn \\
&& + \frac{a_2}2 (z-1)^2+\cdots,
\end{eqnarray}
which corresponds to
\begin{eqnarray}
&&\hat{R}(\phi)=\hat{R}(0)+\frac{\hat{R}''(0)}2 \phi^2+\frac{\hat{R}'''(0^+)}{3!} \phi^3
+\frac{\hat{R}^{(4)}(0)}{4!} \phi^3 +\cdots \nonumber \ \ \ \\
&&
\end{eqnarray}
with
\begin{eqnarray}
&& \hat{R}''(0)=-{R}'(1), \label{eq-RR-1}\\
&& \hat{R}'''(0)=2{a}_1,\label{eq-RR-2}\\
&& \hat{R}^{(4)}(0)={R}'(1)+3{a}_2. \label{eq-RR-3}
\end{eqnarray}

\subsection{Renormalization of the single-replica terms}

From the one-loop correction to the term $(\nabla \bm{\pi}_a)^2$ we find
\begin{eqnarray}
\frac1{\hbar'}= \frac{\zeta^2}{\hbar}
\left[1+ \hbar J_1 + {R}'(1)J_2 \right],
\end{eqnarray}
where the one-loop integrals are given by
\begin{eqnarray}
J_1&=& \frac{K_d T}{\rho \hbar}\sum\limits_{m=-\infty}^{\infty}
\int_{\Lambda e^{-\ell}}^{\Lambda}
\frac{q^{d-1} dq}{q^2+c^{-2} \omega_m^2}  \nn \\
&& =
\frac{K_d}{2 \rho}\, c \Lambda^{d-1} (1-e^{-\ell})
 \coth\Big[\frac{c\hbar\Lambda}{2T} \Big] \label{eq-A-J1}
\end{eqnarray}
and
\begin{eqnarray}
J_2=\frac{K_d}{\rho^2}\int_{\Lambda e^{-\ell}}^{\Lambda} \frac{q^{d-1}dq}{q^4} =
\frac{K_d}{\rho^2} \frac{\Lambda^{d-4}}{d-4}[{1-e^{-\ell (d-4)}}].
\ \ \ \ \ \ \ \ \  \label{eq-A-J2}
\end{eqnarray}
The correction to the external field reads
\begin{eqnarray}
\frac{h'}{\hbar'}= \zeta^2\left(\frac{h}{\hbar}\right)
\left[1+ \frac12(N-1)[\hbar J_1 + {R}'(1)J_2   ] \right]. \label{eq-n1-hbar} \ \ \ \ \ \
\end{eqnarray}
The spin rescaling factor $\zeta$ can be found by noting
that the combination $h/\hbar$ renormalizes trivially
as~\cite{nelson1977momentum}
${h}'/{\hbar}'= \zeta \left({h}/{\hbar}\right)$.
This gives
\begin{eqnarray}
\zeta = 1- \frac12(N-1)[\hbar J_1 + {R}'(1)J_2 ]
\end{eqnarray}
and
\begin{eqnarray}
\hbar' = \hbar[1 +(N-2)(\hbar J_1 + R'(1)J_2 )]. \label{eq-hbar-n}
\end{eqnarray}
The renormalization of  the spin wave velocity can be found from the
correction to the $(\partial_\tau\bm{\pi}_{a})^2$ term:
\begin{eqnarray}
\frac{(\omega_{m}')^2}{c'^2 \hbar'}
&=&\zeta^2 \left(\frac{\omega^2}{c^2 \hbar}\right)
\left[1 + \hbar J_1 + 2 {R}'(1)J_2  \right. \nn \\
&& \left. +(N+1) a_2 J_2  \right],
\end{eqnarray}
which can be rewritten as
\begin{eqnarray}
\frac{c'\hbar'}{T'} &=& \frac{c\hbar}{T} [1- \frac16(N+1)\hat{R}^{(4)}(0)J_2 \nn \\
 && -\frac16(N-2)\hat{R}''(0)J_2]. \label{eq-n1-c}
\end{eqnarray}
Using the definitions (\ref{eq-red-R})-(\ref{eq-red-T}) we derive from
Eqs.~(\ref{eq-n1-hbar})-(\ref{eq-n1-c}) the
flow equations~(\ref{eq-zeta}) and (\ref{eq-flow-T-h}).

\subsection{Correction to disorder}

The disorder term contains two replicas: To find its renormalization
it is convenient to expand around some background state
$\mathbf{n}^0_a(\tau,x)$ which depends explicitly on replica index $a$ and
slowly changes in space.\cite{ledoussal2006random}
For a particular pair of replicas $a$ and $b$
we reparametrize $\mathbf{n}_a(\tau,x)$ and $\mathbf{n}_b(\tau,x)$
as $\mathbf{n}_a(\tau,x)=(\sigma_a,\eta_a,\bm{\rho}_a)$ and
$\mathbf{n}_b(\tau,x)=(\sigma_b,\eta_b,\bm{\rho}_b)$ where  the $\sigma$
and $\eta$ components lie in the plane spanned by vectors $\mathbf{n}^0_a$ and
$\mathbf{n}^0_b$ in a such way that  $\sigma_a=\sqrt{1-\eta_a^2-\bm{\rho}_a^2}$
is parallel to $\mathbf{n}^0_a$
and  $\sigma_b=\sqrt{1-\eta_b^2-\bm{\rho}_b^2}$ to $\mathbf{n}^0_b$.
The components $\bm{\rho}$  are orthogonal to the plane.
Defining the angle between $\mathbf{n}_a^0$
and $\mathbf{n}_b^0$ as $\phi_{ab}$ we obtain
$\mathbf{n}_a\cdot \mathbf{n}_b = \bm{\rho}_a \cdot \bm{\rho}_b +
\cos\phi_{ab}(\sigma_a\sigma_b + \eta_a\eta_b )+ \sin\phi_{ab}
(\sigma_a \eta_b - \sigma_b\eta_a )$.
Expanding in small $\eta$ and $\bm{\rho}$ we get
\begin{eqnarray}
&& \mathbf{n}_a\cdot \mathbf{n}_b = \bm{\rho}_a \cdot \bm{\rho}_b
+
\cos\phi_{ab}[1-\frac12(\eta_a^2+\bm{\rho}_a^2 \nn \\
&& +\eta_b^2+\bm{\rho}_b^2) +\eta_a\eta_b ] + \sin\phi_{ab}\nonumber \\
&& \times
\left[ \eta_b - \eta_a - \frac12 \eta_b(\eta_a^2+\bm{\rho}_a^2)+
 \frac12 \eta_a(\eta_b^2+\bm{\rho}_b^2) \right]. \label{eq-cosphi}
\end{eqnarray}
Substituting Eq. (\ref{eq-cosphi}) into $R$ and expanding again in small
$\eta$ and $\bm{\rho}$ to second order we obtain
\begin{eqnarray}
{R}(\mathbf{n}_a\cdot \mathbf{n}_b)&=&{R}(\cos\phi_{ab})
+ {R}'(\cos\phi_{ab})
\left\{\bm{\rho}_a \cdot \bm{\rho}_b \right.\nn \\
&& \left. -\frac12 \cos\phi_{ab} [
\bm{\rho}_a^2+\bm{\rho}_b^2 + (\eta_a - \eta_b)^2] \right\}
 \nn \\
&& +\frac12 {R}''(\cos\phi_{ab}) (\eta_a - \eta_b)^2 \sin^2\phi_{ab}. \ \ \ \
\label{eq201}
\end{eqnarray}
Using that $R(\cos\phi)=\hat{R}(\phi)$,
${R}'(\cos\phi)=-\hat{R}'(\phi)/\sin\phi$ and
${R}''(\cos\phi)=[\hat{R}''(\phi)-\hat{R}'(\phi)\cos\phi/\sin\phi]/\sin^2\phi$
we can rewrite
Eq.~(\ref{eq201}) as
\begin{eqnarray}
{R}(\mathbf{n}_a\cdot \mathbf{n}_b)&=&\hat{R}(\phi_{ab})
+ \hat{R}'(\phi_{ab})
\left\{
\frac{\bm{\rho}_a^2+\bm{\rho}_b^2} {2\tan\phi_{ab}}
-\frac{\bm{\rho}_a \cdot \bm{\rho}_b}{\sin\phi_{ab}}  \right\}
 \nn \\
&& +\frac12 \hat{R}''(\phi_{ab}) (\eta_a - \eta_b)^2 .
\label{eq20}
\end{eqnarray}
To compute the one-loop correction to the disorder correlator
we expand
$\exp[-\mathcal{S}/\hbar]$ to second order in
${R}(\mathbf{n}_a\cdot \mathbf{n}_b)$ and
perform Gaussian integration over $\eta$ and $\bm\rho$ assuming that they contain
only the fast parts of $\bm\pi_a$.
To extract the one-loop correction it is enough to keep the terms
quartic in $\eta$ and $\bm\rho$, which gives
\begin{eqnarray}
&&\delta^{(1)}\left[\frac{\hat{R}(\phi_{ab})}{2\hbar^2}\right]=
\frac1{8\hbar^4} \sum\limits_{abcd} \Big{\{} \frac14 \hat{R}''(\phi_{ab})
\hat{R}''(\phi_{cd})  \nonumber \\
&& \times \left\langle (\eta_a - \eta_b)^2(\eta_c - \eta_d)^2\right\rangle
+ \hat{R}'(\phi_{ab})\hat{R}'(\phi_{cd}) \nn \\
&&\left \langle\left(\frac{\bm{\rho}_a^2+\bm{\rho}_b^2} {2\tan\phi_{ab}}
-\frac{\bm{\rho}_a \cdot \bm{\rho}_b}{\sin\phi_{ab}}\right)
\left(\frac{\bm{\rho}_c^2+\bm{\rho}_d^2} {2\tan\phi_{cd}}
-\frac{\bm{\rho}_c \cdot \bm{\rho}_d}{\sin\phi_{cd}}\right) \right\rangle
 \Big{\}}. \nn \\ \label{eq-delta1-Gamma}
\end{eqnarray}
Since the integration over $\eta$ and $\bm\rho$ is Gaussian we can use the Wick
theorem with the following contractions:
\begin{eqnarray}
\langle \rho_a^i(q,\omega) \rho_b^j(-q,-\omega)\rangle
&=& \frac{\hbar\, \delta^{ij}\,\delta_{ab}
}{q^2+c_0^{-2} \omega^2+h}, \nn \\
\langle \eta_a(q,\omega) \eta_b(-q,-\omega)\rangle &=& \frac{\hbar\, \delta_{ab}
}{q^2+c_0^{-2} \omega^2+h}.
\end{eqnarray}
Using this in Eq.~(\ref{eq-delta1-Gamma}) we obtain one loop diagrams
in which there is non-zero momentum circulation while
there is no frequency  circulation since the disorder vertices
do not transmit it. For instance we have
\begin{eqnarray}
\left\langle (\eta_a - \eta_b)^2(\eta_c - \eta_d)^2\right\rangle=
8\hbar^2 J_2 \delta_{ac}\delta_{bd} - 16 \hbar^2 J_2 \delta_{ac}\delta_{ad}, \nn
\end{eqnarray}
which can be depicted using the following diagrams
\begin{eqnarray}
{\includegraphics[width=60mm]{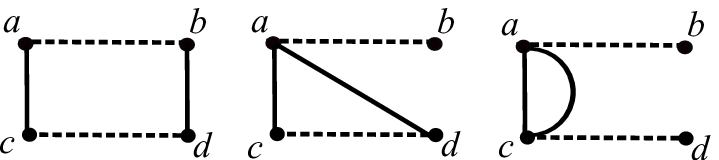}} \nn
\end{eqnarray}
The last diagram is proportional to $\delta_{ac}\delta_{ac}$. After summation
over replica indices it gives the number of replicas,
and thus, vanishes in the
limit of zero replicas $n \to 0$.
Similarly we obtain
\begin{eqnarray}
\left\langle (\bm{\rho}_a \cdot \bm{\rho}_b)(\bm{\rho}_c \cdot \bm{\rho}_d)\right\rangle=
2(N-2)\hbar^2 J_2 \delta_{ac}\delta_{bd}, \nn
\end{eqnarray}
and
\begin{eqnarray}
\frac{\hat{R}'(\phi_{cd})}{\sin\phi_{cd}}\left\langle (\bm{\rho}_a^2 )
(\bm{\rho}_c \cdot \bm{\rho}_d)\right\rangle=
2(N-2)\hbar^2 J_2 \delta_{ac}\delta_{bd} \frac{\hat{R}'(\phi_{aa})}{\sin\phi_{aa}} , \nn
\end{eqnarray}
where we can replace ${\hat{R}'(\phi_{aa})}/{\sin\phi_{aa}}$ by
$\hat{R}''(0)$ using that $\phi_{aa}=0$.
Denoting $\phi:=\phi_{ab}$ we arrive at
\begin{eqnarray}
&&\delta^{(1)}\left[\frac{\hat{R}(\phi)}{2\hbar^2}\right]=
\frac1{2\hbar^2}  \Big{\{} \frac12 [\hat{R}''(\phi)]^2 -\hat{R}''(\phi)\hat{R}''(0)
 \nonumber \\
&& + (N-2)\left(\frac12 \frac{\hat{R}'(\phi)^2}{\sin^2 \phi}-
\frac{\hat{R}'(\phi)\hat{R}''(0)}{\tan \phi}\right)\Big{\}}J_2. \ \
\end{eqnarray}
The correction to the disorder correlator  due to finite $\hbar$ is given by
\begin{eqnarray}
\delta^{(2)}\left[\frac{\hat{R}(\phi)}{2\hbar^2}\right]=
&&\Big \langle -R'(\cos\phi_{ab}) \cos\phi_{ab}
  [\bm{\rho}_b^2+\eta_b^2] \nn \\
&&+
 R''(\cos\phi_{ab})\sin^2\phi_{ab} \eta_b^2 \Big \rangle \nn \\
&& = \hbar J_1 \left[R'(\cos\phi_{ab}) \{-\cos\phi_{ab}
  [(N-2)+1]\} \right. \nn \\
&&+\left.
R''(\cos\phi_{ab})\sin^2\phi_{ab}  \right] \nn \\
&& = \hbar J_1 \Big[(N-2)\frac{\hat{R}'(\phi)}{\tan \phi} + \hat{R}''(\phi)\Big].
\end{eqnarray}
Using
\begin{eqnarray}
\frac{[\hat{R}(\phi)]'}{2\hbar'^2}=\frac{\hat{R}(\phi)}{2\hbar^2} +
\delta^{(1)}\left[\frac{\hat{R}(\phi)}{2\hbar^2}\right]+
\delta^{(2)}\left[\frac{\hat{R}(\phi)}{2\hbar^2}\right]
\end{eqnarray}
and Eq.~(\ref{eq-hbar-n}) we derive
\begin{eqnarray}
&& \left[\hat{R}(\phi)\right]'=
\hat{R}(\phi) + \hat{R}''(\phi)[\hbar J_1 -\hat{R}''(0)J_2]
  \nn \\
&&+ \frac12 [\hat{R}''(\phi)]^2 J_2
 + (N-2)\left(\frac12 \frac{\hat{R}'(\phi)^2}{\sin^2 \phi}J_2 \right. \nn \\
&& + \left.
\left[\frac{\hat{R}'(\phi)}{\tan \phi}+2\hat{R}(\phi)\right][\hbar J_1 - \hat{R}''(0)J_2]
\right), \nn \\ \label{eq-n123n}
\end{eqnarray}
where $J_1$ and $J_2$ are given by Eqs.~(\ref{eq-A-J1}) and (\ref{eq-A-J2}).
The flow equation for the
disorder correlator~(\ref{eq-FRG-1-n}) follows from Eq.~(\ref{eq-n123n}).

\section{Renormalization of the entire spectrum} \label{sec:B}

Renormalization of the generalized action~(\ref{eq-action-rep-3})
resembles renormalization of the clean
NL$\sigma$M with a damping term,~\cite{gamba1999renormalization}
\begin{eqnarray}
D_0(\omega)=\frac{\omega^2}{c^2} + f(\omega).
\end{eqnarray}
The particular form of the damping term $f(\omega)$
depends on the mechanism of damping and it has to be introduced into the clean
NL$\sigma$M by hand.~\cite{gamba1999renormalization}
In our model the form of the spectrum is renormalized by disorder.
Indeed, the one-loop correction to $D(\omega)$ reads
\begin{eqnarray}
 \frac{\left[D(\omega)\right]'}{\hbar'} &= & \frac{\zeta^2}{\hbar}
\left\{ D(\omega) + \hbar I_2(\omega) + {R}'(1) [D(\omega)J_2 + I_1(\omega)] \right. \notag \\
&& \left. + {a}_2 (N+1) I_1(\omega) \right\}, \label{eq-DDD}
\end{eqnarray}
where
\begin{eqnarray}
I_1(\omega)&=&  \frac1{\rho^2}\int_q \left[ \frac1{q^2+h}- \frac1{q^2+D(\omega)+h} \right],  \\
I_2(\omega)&=& \frac1{\rho} \sum\limits_{m=-\infty}^{\infty} \int_q
    \frac{D(\omega+\omega_m)-D(\omega_m)}{q^2+D(\omega_m)+h}.
\end{eqnarray}
Using Eqs.~(\ref{eq-hbar-n}) and (\ref{eq-DDD}) we obtain
\begin{eqnarray}
[D(\omega)]' &=&
 D(\omega) + \hbar[I_2(\omega)-D(\omega)J_1] \nn \\
&& + [R'(1)+a_2 (N+1)] I_1(\omega).
\end{eqnarray}
The first term $\hbar[I_2(\omega)-D(\omega)J_1]$ which was found for the pure
NL$\sigma$M in Ref.~\onlinecite{gamba1999renormalization} is irrelevant at the
zero temperature quasi-classical FP.
The integral in the second term gives
\begin{eqnarray}
I_1(\omega)&=& \frac{K_d}{\rho^2} \int_{\Lambda e^{-\ell}}^{\Lambda}
 \frac{D(\omega)q^{d-1} dq}{q^2[q^2+D(\omega)]} \nn \\
 && = \frac{K_d}{\rho^2} \frac{D(\omega)\Lambda^{d-2} \ell}{\Lambda^2+D(\omega)}.
\end{eqnarray}
Introducing
$\tilde{D}(\omega) = \Lambda_\ell^{-2} D(\omega)$ and using
Eq.~(\ref{eq-red-R})
we find the flow equation for the spectrum as
\begin{eqnarray}
\partial_\ell \tilde{D}(\omega) &=& 2 \tilde{D}(\omega)
+\frac13[(N+1)\tilde{R}^{(4)}(0) \nn \\
&& +(N-2)\tilde{R}^{\prime\prime}(0)] \frac{\tilde{D}(\omega)}{1+\tilde{D}(\omega)}.
\label{eq-D-2}
\end{eqnarray}
Retaining  in Eq.~(\ref{eq-D-2}) only the terms which are dominant in the
vicinity of a zero temperature quasi-classical FP we arrive at Eq.~(\ref{eq-D-1}).


\begin{thebibliography}{45}%
\makeatletter
\providecommand \@ifxundefined [1]{%
 \@ifx{#1\undefined}
}%
\providecommand \@ifnum [1]{%
 \ifnum #1\expandafter \@firstoftwo
 \else \expandafter \@secondoftwo
 \fi
}%
\providecommand \@ifx [1]{%
 \ifx #1\expandafter \@firstoftwo
 \else \expandafter \@secondoftwo
 \fi
}%
\providecommand \natexlab [1]{#1}%
\providecommand \enquote  [1]{``#1''}%
\providecommand \bibnamefont  [1]{#1}%
\providecommand \bibfnamefont [1]{#1}%
\providecommand \citenamefont [1]{#1}%
\providecommand \href@noop [0]{\@secondoftwo}%
\providecommand \href [0]{\begingroup \@sanitize@url \@href}%
\providecommand \@href[1]{\@@startlink{#1}\@@href}%
\providecommand \@@href[1]{\endgroup#1\@@endlink}%
\providecommand \@sanitize@url [0]{\catcode `\\12\catcode `\$12\catcode
  `\&12\catcode `\#12\catcode `\^12\catcode `\_12\catcode `\%12\relax}%
\providecommand \@@startlink[1]{}%
\providecommand \@@endlink[0]{}%
\providecommand \url  [0]{\begingroup\@sanitize@url \@url }%
\providecommand \@url [1]{\endgroup\@href {#1}{\urlprefix }}%
\providecommand \urlprefix  [0]{URL }%
\providecommand \Eprint [0]{\href }%
\providecommand \doibase [0]{http://dx.doi.org/}%
\providecommand \selectlanguage [0]{\@gobble}%
\providecommand \bibinfo  [0]{\@secondoftwo}%
\providecommand \bibfield  [0]{\@secondoftwo}%
\providecommand \translation [1]{[#1]}%
\providecommand \BibitemOpen [0]{}%
\providecommand \bibitemStop [0]{}%
\providecommand \bibitemNoStop [0]{.\EOS\space}%
\providecommand \EOS [0]{\spacefactor3000\relax}%
\providecommand \BibitemShut  [1]{\csname bibitem#1\endcsname}%
\let\auto@bib@innerbib\@empty
\bibitem [{\citenamefont {Evers}\ and\ \citenamefont
  {Mirlin}(2008)}]{evers2008anderson}%
  \BibitemOpen
  \bibfield  {author} {\bibinfo {author} {\bibfnamefont {F.}~\bibnamefont
  {Evers}}\ and\ \bibinfo {author} {\bibfnamefont {A.~D.}\ \bibnamefont
  {Mirlin}},\ }\href {\doibase 10.1103/RevModPhys.80.1355} {\bibfield
  {journal} {\bibinfo  {journal} {Rev. Mod. Phys.}\ }\textbf {\bibinfo {volume}
  {80}},\ \bibinfo {pages} {1355} (\bibinfo {year} {2008})}\BibitemShut
  {NoStop}%
\bibitem [{\citenamefont {Basko}\ \emph {et~al.}(2006)\citenamefont {Basko},
  \citenamefont {Aleiner},\ and\ \citenamefont {Altshuler}}]{basko2006metal}%
  \BibitemOpen
  \bibfield  {author} {\bibinfo {author} {\bibfnamefont {D.~M.}\ \bibnamefont
  {Basko}}, \bibinfo {author} {\bibfnamefont {I.~L.}\ \bibnamefont {Aleiner}},
  \ and\ \bibinfo {author} {\bibfnamefont {B.~L.}\ \bibnamefont {Altshuler}},\
  }\href@noop {} {\bibfield  {journal} {\bibinfo  {journal} {Annals of
  Physics}\ }\textbf {\bibinfo {volume} {321}},\ \bibinfo {pages} {1126}
  (\bibinfo {year} {2006})}\BibitemShut {NoStop}%
\bibitem [{\citenamefont {Aleiner}\ \emph {et~al.}(2010)\citenamefont
  {Aleiner}, \citenamefont {Altshuler},\ and\ \citenamefont
  {Shlyapnikov}}]{aleiner2010finite}%
  \BibitemOpen
  \bibfield  {author} {\bibinfo {author} {\bibfnamefont {I.~L.}\ \bibnamefont
  {Aleiner}}, \bibinfo {author} {\bibfnamefont {B.~L.}\ \bibnamefont
  {Altshuler}}, \ and\ \bibinfo {author} {\bibfnamefont {G.~V.}\ \bibnamefont
  {Shlyapnikov}},\ }\href@noop {} {\bibfield  {journal} {\bibinfo  {journal}
  {Nature Physics}\ }\textbf {\bibinfo {volume} {6}},\ \bibinfo {pages} {900}
  (\bibinfo {year} {2010})}\BibitemShut {NoStop}%
\bibitem [{\citenamefont {Yu}\ and\ \citenamefont
  {Mueller}(2013)}]{Yu2013localization}%
  \BibitemOpen
  \bibfield  {author} {\bibinfo {author} {\bibfnamefont {X.}~\bibnamefont
  {Yu}}\ and\ \bibinfo {author} {\bibfnamefont {M.}~\bibnamefont {Mueller}},\
  }\href@noop {} {\bibfield  {journal} {\bibinfo  {journal} {Annals of
  Physics}\ }\textbf {\bibinfo {volume} {337}},\ \bibinfo {pages} {55 }
  (\bibinfo {year} {2013})}\BibitemShut {NoStop}%
\bibitem [{\citenamefont {Pal}\ and\ \citenamefont {Huse}(2010)}]{pal2010many}%
  \BibitemOpen
  \bibfield  {author} {\bibinfo {author} {\bibfnamefont {A.}~\bibnamefont
  {Pal}}\ and\ \bibinfo {author} {\bibfnamefont {D.~A.}\ \bibnamefont {Huse}},\
  }\href {\doibase 10.1103/PhysRevB.82.174411} {\bibfield  {journal} {\bibinfo
  {journal} {Phys. Rev. B}\ }\textbf {\bibinfo {volume} {82}},\ \bibinfo
  {pages} {174411} (\bibinfo {year} {2010})}\BibitemShut {NoStop}%
\bibitem [{\citenamefont {Vosk}\ and\ \citenamefont
  {Altman}(2013)}]{vosk2013many}%
  \BibitemOpen
  \bibfield  {author} {\bibinfo {author} {\bibfnamefont {R.}~\bibnamefont
  {Vosk}}\ and\ \bibinfo {author} {\bibfnamefont {E.}~\bibnamefont {Altman}},\
  }\href {\doibase 10.1103/PhysRevLett.110.067204} {\bibfield  {journal}
  {\bibinfo  {journal} {Phys. Rev. Lett.}\ }\textbf {\bibinfo {volume} {110}},\
  \bibinfo {pages} {067204} (\bibinfo {year} {2013})}\BibitemShut {NoStop}%
\bibitem [{\citenamefont {Fisher}\ \emph {et~al.}(1989)\citenamefont {Fisher},
  \citenamefont {Weichman}, \citenamefont {Grinstein},\ and\ \citenamefont
  {Fisher}}]{fisher1989boson}%
  \BibitemOpen
  \bibfield  {author} {\bibinfo {author} {\bibfnamefont {M.~P.~A.}\
  \bibnamefont {Fisher}}, \bibinfo {author} {\bibfnamefont {P.~B.}\
  \bibnamefont {Weichman}}, \bibinfo {author} {\bibfnamefont {G.}~\bibnamefont
  {Grinstein}}, \ and\ \bibinfo {author} {\bibfnamefont {D.~S.}\ \bibnamefont
  {Fisher}},\ }\href {\doibase 10.1103/PhysRevB.40.546} {\bibfield  {journal}
  {\bibinfo  {journal} {Phys. Rev. B}\ }\textbf {\bibinfo {volume} {40}},\
  \bibinfo {pages} {546} (\bibinfo {year} {1989})}\BibitemShut {NoStop}%
\bibitem [{\citenamefont {Nelson}\ and\ \citenamefont
  {Vinokur}(1992)}]{nelson1992boson}%
  \BibitemOpen
  \bibfield  {author} {\bibinfo {author} {\bibfnamefont {D.~R.}\ \bibnamefont
  {Nelson}}\ and\ \bibinfo {author} {\bibfnamefont {V.~M.}\ \bibnamefont
  {Vinokur}},\ }\href {\doibase 10.1103/PhysRevLett.68.2398} {\bibfield
  {journal} {\bibinfo  {journal} {Phys. Rev. Lett.}\ }\textbf {\bibinfo
  {volume} {68}},\ \bibinfo {pages} {2398} (\bibinfo {year}
  {1992})}\BibitemShut {NoStop}%
\bibitem [{\citenamefont {Giamarchi}\ and\ \citenamefont
  {Le~Doussal}(1996)}]{giamarchi1996variational}%
  \BibitemOpen
  \bibfield  {author} {\bibinfo {author} {\bibfnamefont {T.}~\bibnamefont
  {Giamarchi}}\ and\ \bibinfo {author} {\bibfnamefont {P.}~\bibnamefont
  {Le~Doussal}},\ }\href {\doibase 10.1103/PhysRevB.53.15206} {\bibfield
  {journal} {\bibinfo  {journal} {Phys. Rev. B}\ }\textbf {\bibinfo {volume}
  {53}},\ \bibinfo {pages} {15206} (\bibinfo {year} {1996})}\BibitemShut
  {NoStop}%
\bibitem [{\citenamefont {Fedorenko}(2008)}]{fedorenko2008elastic}%
  \BibitemOpen
  \bibfield  {author} {\bibinfo {author} {\bibfnamefont {A.~A.}\ \bibnamefont
  {Fedorenko}},\ }\href {\doibase 10.1103/PhysRevB.77.094203} {\bibfield
  {journal} {\bibinfo  {journal} {Phys. Rev. B}\ }\textbf {\bibinfo {volume}
  {77}},\ \bibinfo {pages} {094203} (\bibinfo {year} {2008})}\BibitemShut
  {NoStop}%
\bibitem [{\citenamefont {Yu}\ \emph {et~al.}(2012)\citenamefont {Yu},
  \citenamefont {Yin}, \citenamefont {Sullivan}, \citenamefont {Xia},
  \citenamefont {Huan}, \citenamefont {Paduan-Filho}, \citenamefont {Oliveira},
  \citenamefont {Haas}, \citenamefont {Steppke}, \citenamefont {Miclea},
  \citenamefont {Weickert}, \citenamefont {Movshovich}, \citenamefont {Mun},
  \citenamefont {Scott}, \citenamefont {Zapf},\ and\ \citenamefont
  {Roscilde}}]{yu2012bose}%
  \BibitemOpen
  \bibfield  {author} {\bibinfo {author} {\bibfnamefont {R.}~\bibnamefont
  {Yu}}, \bibinfo {author} {\bibfnamefont {L.}~\bibnamefont {Yin}}, \bibinfo
  {author} {\bibfnamefont {N.~S.}\ \bibnamefont {Sullivan}}, \bibinfo {author}
  {\bibfnamefont {J.~S.}\ \bibnamefont {Xia}}, \bibinfo {author} {\bibfnamefont
  {C.}~\bibnamefont {Huan}}, \bibinfo {author} {\bibfnamefont {A.}~\bibnamefont
  {Paduan-Filho}}, \bibinfo {author} {\bibfnamefont {N.~F.}\ \bibnamefont
  {Oliveira}, \bibfnamefont {Jr.}}, \bibinfo {author} {\bibfnamefont
  {S.}~\bibnamefont {Haas}}, \bibinfo {author} {\bibfnamefont {A.}~\bibnamefont
  {Steppke}}, \bibinfo {author} {\bibfnamefont {C.~F.}\ \bibnamefont {Miclea}},
  \bibinfo {author} {\bibfnamefont {F.}~\bibnamefont {Weickert}}, \bibinfo
  {author} {\bibfnamefont {R.}~\bibnamefont {Movshovich}}, \bibinfo {author}
  {\bibfnamefont {E.-D.}\ \bibnamefont {Mun}}, \bibinfo {author} {\bibfnamefont
  {B.~L.}\ \bibnamefont {Scott}}, \bibinfo {author} {\bibfnamefont {V.~S.}\
  \bibnamefont {Zapf}}, \ and\ \bibinfo {author} {\bibfnamefont
  {T.}~\bibnamefont {Roscilde}},\ }\href@noop {} {\bibfield  {journal}
  {\bibinfo  {journal} {Nature}\ }\textbf {\bibinfo {volume} {489}},\ \bibinfo
  {pages} {379} (\bibinfo {year} {2012})}\BibitemShut {NoStop}%
\bibitem [{\citenamefont {Orignac}\ \emph {et~al.}(1999)\citenamefont
  {Orignac}, \citenamefont {Giamarchi},\ and\ \citenamefont
  {Le~Doussal}}]{orignac1999possible}%
  \BibitemOpen
  \bibfield  {author} {\bibinfo {author} {\bibfnamefont {E.}~\bibnamefont
  {Orignac}}, \bibinfo {author} {\bibfnamefont {T.}~\bibnamefont {Giamarchi}},
  \ and\ \bibinfo {author} {\bibfnamefont {P.}~\bibnamefont {Le~Doussal}},\
  }\href {\doibase 10.1103/PhysRevLett.83.2378} {\bibfield  {journal} {\bibinfo
   {journal} {Phys. Rev. Lett.}\ }\textbf {\bibinfo {volume} {83}},\ \bibinfo
  {pages} {2378} (\bibinfo {year} {1999})}\BibitemShut {NoStop}%
\bibitem [{\citenamefont {Giamarchi}\ \emph {et~al.}(2001)\citenamefont
  {Giamarchi}, \citenamefont {Le~Doussal},\ and\ \citenamefont
  {Orignac}}]{giamarchi2001competition}%
  \BibitemOpen
  \bibfield  {author} {\bibinfo {author} {\bibfnamefont {T.}~\bibnamefont
  {Giamarchi}}, \bibinfo {author} {\bibfnamefont {P.}~\bibnamefont
  {Le~Doussal}}, \ and\ \bibinfo {author} {\bibfnamefont {E.}~\bibnamefont
  {Orignac}},\ }\href {\doibase 10.1103/PhysRevB.64.245119} {\bibfield
  {journal} {\bibinfo  {journal} {Phys. Rev. B}\ }\textbf {\bibinfo {volume}
  {64}},\ \bibinfo {pages} {245119} (\bibinfo {year} {2001})}\BibitemShut
  {NoStop}%
\bibitem [{\citenamefont {Z\'u\~niga}\ and\ \citenamefont
  {Laflorencie}(2013)}]{zuninga2013bose}%
  \BibitemOpen
  \bibfield  {author} {\bibinfo {author} {\bibfnamefont {J.~P.~A.}\
  \bibnamefont {Z\'u\~niga}}\ and\ \bibinfo {author} {\bibfnamefont
  {N.}~\bibnamefont {Laflorencie}},\ }\href {\doibase
  10.1103/PhysRevLett.111.160403} {\bibfield  {journal} {\bibinfo  {journal}
  {Phys. Rev. Lett.}\ }\textbf {\bibinfo {volume} {111}},\ \bibinfo {pages}
  {160403} (\bibinfo {year} {2013})}\BibitemShut {NoStop}%
\bibitem [{\citenamefont {Gangopadhyay}\ \emph {et~al.}(2013)\citenamefont
  {Gangopadhyay}, \citenamefont {Galitski},\ and\ \citenamefont
  {M\"uller}}]{gangopadhyay2013magnetoresistance}%
  \BibitemOpen
  \bibfield  {author} {\bibinfo {author} {\bibfnamefont {A.}~\bibnamefont
  {Gangopadhyay}}, \bibinfo {author} {\bibfnamefont {V.}~\bibnamefont
  {Galitski}}, \ and\ \bibinfo {author} {\bibfnamefont {M.}~\bibnamefont
  {M\"uller}},\ }\href {\doibase 10.1103/PhysRevLett.111.026801} {\bibfield
  {journal} {\bibinfo  {journal} {Phys. Rev. Lett.}\ }\textbf {\bibinfo
  {volume} {111}},\ \bibinfo {pages} {026801} (\bibinfo {year}
  {2013})}\BibitemShut {NoStop}%
\bibitem [{\citenamefont {Fisher}(1986{\natexlab{a}})}]{fisher1986interface}%
  \BibitemOpen
  \bibfield  {author} {\bibinfo {author} {\bibfnamefont {D.~S.}\ \bibnamefont
  {Fisher}},\ }\href {\doibase 10.1103/PhysRevLett.56.1964} {\bibfield
  {journal} {\bibinfo  {journal} {Phys. Rev. Lett.}\ }\textbf {\bibinfo
  {volume} {56}},\ \bibinfo {pages} {1964} (\bibinfo {year}
  {1986}{\natexlab{a}})}\BibitemShut {NoStop}%
\bibitem [{\citenamefont {Nattermann}\ \emph {et~al.}(1992)\citenamefont
  {Nattermann}, \citenamefont {Stepanow}, \citenamefont {Tang},\ and\
  \citenamefont {Leschhorn}}]{nattermann1992dyanmics}%
  \BibitemOpen
  \bibfield  {author} {\bibinfo {author} {\bibfnamefont {T.}~\bibnamefont
  {Nattermann}}, \bibinfo {author} {\bibfnamefont {S.}~\bibnamefont
  {Stepanow}}, \bibinfo {author} {\bibfnamefont {L.-H.}\ \bibnamefont {Tang}},
  \ and\ \bibinfo {author} {\bibfnamefont {H.}~\bibnamefont {Leschhorn}},\
  }\href {\doibase 10.1051/jp2:1992214} {\bibfield  {journal} {\bibinfo
  {journal} {J. Phys. II France}\ }\textbf {\bibinfo {volume} {2}},\ \bibinfo
  {pages} {1483} (\bibinfo {year} {1992})}\BibitemShut {NoStop}%
\bibitem [{\citenamefont {Chauve}\ \emph {et~al.}(2001)\citenamefont {Chauve},
  \citenamefont {Le~Doussal},\ and\ \citenamefont
  {Wiese}}]{chauve2001renormalization}%
  \BibitemOpen
  \bibfield  {author} {\bibinfo {author} {\bibfnamefont {P.}~\bibnamefont
  {Chauve}}, \bibinfo {author} {\bibfnamefont {P.}~\bibnamefont {Le~Doussal}},
  \ and\ \bibinfo {author} {\bibfnamefont {K.~J.}\ \bibnamefont {Wiese}},\
  }\href {\doibase 10.1103/PhysRevLett.86.1785} {\bibfield  {journal} {\bibinfo
   {journal} {Phys. Rev. Lett.}\ }\textbf {\bibinfo {volume} {86}},\ \bibinfo
  {pages} {1785} (\bibinfo {year} {2001})}\BibitemShut {NoStop}%
\bibitem [{\citenamefont {Fedorenko}\ \emph {et~al.}(2006)\citenamefont
  {Fedorenko}, \citenamefont {Le~Doussal},\ and\ \citenamefont
  {Wiese}}]{fedorenko2006statics}%
  \BibitemOpen
  \bibfield  {author} {\bibinfo {author} {\bibfnamefont {A.~A.}\ \bibnamefont
  {Fedorenko}}, \bibinfo {author} {\bibfnamefont {P.}~\bibnamefont
  {Le~Doussal}}, \ and\ \bibinfo {author} {\bibfnamefont {K.~J.}\ \bibnamefont
  {Wiese}},\ }\href {\doibase 10.1103/PhysRevE.74.061109} {\bibfield  {journal}
  {\bibinfo  {journal} {Phys. Rev. E}\ }\textbf {\bibinfo {volume} {74}},\
  \bibinfo {pages} {061109} (\bibinfo {year} {2006})}\BibitemShut {NoStop}%
\bibitem [{\citenamefont {Fisher}(1986{\natexlab{b}})}]{fisher1986scaling}%
  \BibitemOpen
  \bibfield  {author} {\bibinfo {author} {\bibfnamefont {D.~S.}\ \bibnamefont
  {Fisher}},\ }\href {\doibase 10.1103/PhysRevLett.56.416} {\bibfield
  {journal} {\bibinfo  {journal} {Phys. Rev. Lett.}\ }\textbf {\bibinfo
  {volume} {56}},\ \bibinfo {pages} {416} (\bibinfo {year}
  {1986}{\natexlab{b}})}\BibitemShut {NoStop}%
\bibitem [{\citenamefont {Dutta}\ and\ \citenamefont
  {Bhattacharjee}(1998)}]{dutta1998quantum}%
  \BibitemOpen
  \bibfield  {author} {\bibinfo {author} {\bibfnamefont {A.}~\bibnamefont
  {Dutta}}\ and\ \bibinfo {author} {\bibfnamefont {J.~K.}\ \bibnamefont
  {Bhattacharjee}},\ }\href {\doibase 10.1103/PhysRevB.58.6378} {\bibfield
  {journal} {\bibinfo  {journal} {Phys. Rev. B}\ }\textbf {\bibinfo {volume}
  {58}},\ \bibinfo {pages} {6378} (\bibinfo {year} {1998})}\BibitemShut
  {NoStop}%
\bibitem [{\citenamefont {Senthil}(1998)}]{senthil1998properties}%
  \BibitemOpen
  \bibfield  {author} {\bibinfo {author} {\bibfnamefont {T.}~\bibnamefont
  {Senthil}},\ }\href {\doibase 10.1103/PhysRevB.57.8375} {\bibfield  {journal}
  {\bibinfo  {journal} {Phys. Rev. B}\ }\textbf {\bibinfo {volume} {57}},\
  \bibinfo {pages} {8375} (\bibinfo {year} {1998})}\BibitemShut {NoStop}%
\bibitem [{\citenamefont {Chauve}\ \emph {et~al.}(2000)\citenamefont {Chauve},
  \citenamefont {Giamarchi},\ and\ \citenamefont
  {Le~Doussal}}]{chauve2000creep}%
  \BibitemOpen
  \bibfield  {author} {\bibinfo {author} {\bibfnamefont {P.}~\bibnamefont
  {Chauve}}, \bibinfo {author} {\bibfnamefont {T.}~\bibnamefont {Giamarchi}}, \
  and\ \bibinfo {author} {\bibfnamefont {P.}~\bibnamefont {Le~Doussal}},\
  }\href {\doibase 10.1103/PhysRevB.62.6241} {\bibfield  {journal} {\bibinfo
  {journal} {Phys. Rev. B}\ }\textbf {\bibinfo {volume} {62}},\ \bibinfo
  {pages} {6241} (\bibinfo {year} {2000})}\BibitemShut {NoStop}%
\bibitem [{\citenamefont {Gorokhov}\ \emph {et~al.}(2002)\citenamefont
  {Gorokhov}, \citenamefont {Fisher},\ and\ \citenamefont
  {Blatter}}]{gorokhov2002creep}%
  \BibitemOpen
  \bibfield  {author} {\bibinfo {author} {\bibfnamefont {D.~A.}\ \bibnamefont
  {Gorokhov}}, \bibinfo {author} {\bibfnamefont {D.~S.}\ \bibnamefont
  {Fisher}}, \ and\ \bibinfo {author} {\bibfnamefont {G.}~\bibnamefont
  {Blatter}},\ }\href {\doibase 10.1103/PhysRevB.66.214203} {\bibfield
  {journal} {\bibinfo  {journal} {Phys. Rev. B}\ }\textbf {\bibinfo {volume}
  {66}},\ \bibinfo {pages} {214203} (\bibinfo {year} {2002})}\BibitemShut
  {NoStop}%
\bibitem [{\citenamefont {Ye}\ \emph {et~al.}(1993)\citenamefont {Ye},
  \citenamefont {Sachdev},\ and\ \citenamefont {Read}}]{ye1993solvable}%
  \BibitemOpen
  \bibfield  {author} {\bibinfo {author} {\bibfnamefont {J.}~\bibnamefont
  {Ye}}, \bibinfo {author} {\bibfnamefont {S.}~\bibnamefont {Sachdev}}, \ and\
  \bibinfo {author} {\bibfnamefont {N.}~\bibnamefont {Read}},\ }\href {\doibase
  10.1103/PhysRevLett.70.4011} {\bibfield  {journal} {\bibinfo  {journal}
  {Phys. Rev. Lett.}\ }\textbf {\bibinfo {volume} {70}},\ \bibinfo {pages}
  {4011} (\bibinfo {year} {1993})}\BibitemShut {NoStop}%
\bibitem [{\citenamefont {Andreanov}\ and\ \citenamefont
  {M\"uller}(2012)}]{andreanov2012longrange}%
  \BibitemOpen
  \bibfield  {author} {\bibinfo {author} {\bibfnamefont {A.}~\bibnamefont
  {Andreanov}}\ and\ \bibinfo {author} {\bibfnamefont {M.}~\bibnamefont
  {M\"uller}},\ }\href {\doibase 10.1103/PhysRevLett.109.177201} {\bibfield
  {journal} {\bibinfo  {journal} {Phys. Rev. Lett.}\ }\textbf {\bibinfo
  {volume} {109}},\ \bibinfo {pages} {177201} (\bibinfo {year}
  {2012})}\BibitemShut {NoStop}%
\bibitem [{\citenamefont {Cazalilla}\ \emph {et~al.}(2011)\citenamefont
  {Cazalilla}, \citenamefont {Citro}, \citenamefont {Giamarchi}, \citenamefont
  {Orignac},\ and\ \citenamefont {Rigol}}]{cazalilla2011one}%
  \BibitemOpen
  \bibfield  {author} {\bibinfo {author} {\bibfnamefont {M.~A.}\ \bibnamefont
  {Cazalilla}}, \bibinfo {author} {\bibfnamefont {R.}~\bibnamefont {Citro}},
  \bibinfo {author} {\bibfnamefont {T.}~\bibnamefont {Giamarchi}}, \bibinfo
  {author} {\bibfnamefont {E.}~\bibnamefont {Orignac}}, \ and\ \bibinfo
  {author} {\bibfnamefont {M.}~\bibnamefont {Rigol}},\ }\href {\doibase
  10.1103/RevModPhys.83.1405} {\bibfield  {journal} {\bibinfo  {journal} {Rev.
  Mod. Phys.}\ }\textbf {\bibinfo {volume} {83}},\ \bibinfo {pages} {1405}
  (\bibinfo {year} {2011})}\BibitemShut {NoStop}%
\bibitem [{\citenamefont {Chakravarty}\ \emph {et~al.}(1989)\citenamefont
  {Chakravarty}, \citenamefont {Halperin},\ and\ \citenamefont
  {Nelson}}]{chakravarty1989two}%
  \BibitemOpen
  \bibfield  {author} {\bibinfo {author} {\bibfnamefont {S.}~\bibnamefont
  {Chakravarty}}, \bibinfo {author} {\bibfnamefont {B.~I.}\ \bibnamefont
  {Halperin}}, \ and\ \bibinfo {author} {\bibfnamefont {D.~R.}\ \bibnamefont
  {Nelson}},\ }\href {\doibase 10.1103/PhysRevB.39.2344} {\bibfield  {journal}
  {\bibinfo  {journal} {Phys. Rev. B}\ }\textbf {\bibinfo {volume} {39}},\
  \bibinfo {pages} {2344} (\bibinfo {year} {1989})}\BibitemShut {NoStop}%
\bibitem [{\citenamefont {Demler}\ \emph {et~al.}(2004)\citenamefont {Demler},
  \citenamefont {Hanke},\ and\ \citenamefont {Zhang}}]{demler2004so5}%
  \BibitemOpen
  \bibfield  {author} {\bibinfo {author} {\bibfnamefont {E.}~\bibnamefont
  {Demler}}, \bibinfo {author} {\bibfnamefont {W.}~\bibnamefont {Hanke}}, \
  and\ \bibinfo {author} {\bibfnamefont {S.-C.}\ \bibnamefont {Zhang}},\ }\href
  {\doibase 10.1103/RevModPhys.76.909} {\bibfield  {journal} {\bibinfo
  {journal} {Rev. Mod. Phys.}\ }\textbf {\bibinfo {volume} {76}},\ \bibinfo
  {pages} {909} (\bibinfo {year} {2004})}\BibitemShut {NoStop}%
\bibitem [{\citenamefont {Fisher}(1985)}]{fisher1985random}%
  \BibitemOpen
  \bibfield  {author} {\bibinfo {author} {\bibfnamefont {D.~S.}\ \bibnamefont
  {Fisher}},\ }\href {\doibase 10.1103/PhysRevB.31.7233} {\bibfield  {journal}
  {\bibinfo  {journal} {Phys. Rev. B}\ }\textbf {\bibinfo {volume} {31}},\
  \bibinfo {pages} {7233} (\bibinfo {year} {1985})}\BibitemShut {NoStop}%
\bibitem [{\citenamefont {Fedorenko}\ and\ \citenamefont
  {K\"uhnel}(2007)}]{fedorenko2007long-range}%
  \BibitemOpen
  \bibfield  {author} {\bibinfo {author} {\bibfnamefont {A.~A.}\ \bibnamefont
  {Fedorenko}}\ and\ \bibinfo {author} {\bibfnamefont {F.}~\bibnamefont
  {K\"uhnel}},\ }\href {\doibase 10.1103/PhysRevB.75.174206} {\bibfield
  {journal} {\bibinfo  {journal} {Phys. Rev. B}\ }\textbf {\bibinfo {volume}
  {75}},\ \bibinfo {pages} {174206} (\bibinfo {year} {2007})}\BibitemShut
  {NoStop}%
\bibitem [{\citenamefont {Feldman}(2000)}]{feldman2000quasi}%
  \BibitemOpen
  \bibfield  {author} {\bibinfo {author} {\bibfnamefont {D.~E.}\ \bibnamefont
  {Feldman}},\ }\href {\doibase 10.1103/PhysRevB.61.382} {\bibfield  {journal}
  {\bibinfo  {journal} {Phys. Rev. B}\ }\textbf {\bibinfo {volume} {61}},\
  \bibinfo {pages} {382} (\bibinfo {year} {2000})}\BibitemShut {NoStop}%
\bibitem [{\citenamefont {Proctor}\ \emph {et~al.}(2014)\citenamefont
  {Proctor}, \citenamefont {Garanin},\ and\ \citenamefont
  {Chudnovsky}}]{proctor2014random}%
  \BibitemOpen
  \bibfield  {author} {\bibinfo {author} {\bibfnamefont {T.~C.}\ \bibnamefont
  {Proctor}}, \bibinfo {author} {\bibfnamefont {D.~A.}\ \bibnamefont
  {Garanin}}, \ and\ \bibinfo {author} {\bibfnamefont {E.~M.}\ \bibnamefont
  {Chudnovsky}},\ }\href {\doibase 10.1103/PhysRevLett.112.097201} {\bibfield
  {journal} {\bibinfo  {journal} {Phys. Rev. Lett.}\ }\textbf {\bibinfo
  {volume} {112}},\ \bibinfo {pages} {097201} (\bibinfo {year}
  {2014})}\BibitemShut {NoStop}%
\bibitem [{\citenamefont {Tissier}\ and\ \citenamefont
  {Tarjus}(2006)}]{tissier2006two}%
  \BibitemOpen
  \bibfield  {author} {\bibinfo {author} {\bibfnamefont {M.}~\bibnamefont
  {Tissier}}\ and\ \bibinfo {author} {\bibfnamefont {G.}~\bibnamefont
  {Tarjus}},\ }\href {\doibase 10.1103/PhysRevB.74.214419} {\bibfield
  {journal} {\bibinfo  {journal} {Phys. Rev. B}\ }\textbf {\bibinfo {volume}
  {74}},\ \bibinfo {pages} {214419} (\bibinfo {year} {2006})}\BibitemShut
  {NoStop}%
\bibitem [{\citenamefont {Le~Doussal}\ and\ \citenamefont
  {Wiese}(2006)}]{ledoussal2006random}%
  \BibitemOpen
  \bibfield  {author} {\bibinfo {author} {\bibfnamefont {P.}~\bibnamefont
  {Le~Doussal}}\ and\ \bibinfo {author} {\bibfnamefont {K.~J.}\ \bibnamefont
  {Wiese}},\ }\href {\doibase 10.1103/PhysRevLett.96.197202} {\bibfield
  {journal} {\bibinfo  {journal} {Phys. Rev. Lett.}\ }\textbf {\bibinfo
  {volume} {96}},\ \bibinfo {pages} {197202} (\bibinfo {year}
  {2006})}\BibitemShut {NoStop}%
\bibitem [{\citenamefont {Fedorenko}(2012)}]{fedorenko2012random}%
  \BibitemOpen
  \bibfield  {author} {\bibinfo {author} {\bibfnamefont {A.~A.}\ \bibnamefont
  {Fedorenko}},\ }\href {\doibase 10.1103/PhysRevE.86.021131} {\bibfield
  {journal} {\bibinfo  {journal} {Phys. Rev. E}\ }\textbf {\bibinfo {volume}
  {86}},\ \bibinfo {pages} {021131} (\bibinfo {year} {2012})}\BibitemShut
  {NoStop}%
\bibitem [{\citenamefont {Young}\ and\ \citenamefont
  {Rieger}(1996)}]{young1996numerical}%
  \BibitemOpen
  \bibfield  {author} {\bibinfo {author} {\bibfnamefont {A.~P.}\ \bibnamefont
  {Young}}\ and\ \bibinfo {author} {\bibfnamefont {H.}~\bibnamefont {Rieger}},\
  }\href {\doibase 10.1103/PhysRevB.53.8486} {\bibfield  {journal} {\bibinfo
  {journal} {Phys. Rev. B}\ }\textbf {\bibinfo {volume} {53}},\ \bibinfo
  {pages} {8486} (\bibinfo {year} {1996})}\BibitemShut {NoStop}%
\bibitem [{\citenamefont {Monthus}\ and\ \citenamefont
  {Garel}(2012)}]{monthus2012random}%
  \BibitemOpen
  \bibfield  {author} {\bibinfo {author} {\bibfnamefont {C.}~\bibnamefont
  {Monthus}}\ and\ \bibinfo {author} {\bibfnamefont {T.}~\bibnamefont
  {Garel}},\ }\href {http://stacks.iop.org/1751-8121/45/i=9/a=095002}
  {\bibfield  {journal} {\bibinfo  {journal} {Journal of Physics A:
  Mathematical and Theoretical}\ }\textbf {\bibinfo {volume} {45}},\ \bibinfo
  {pages} {095002} (\bibinfo {year} {2012})}\BibitemShut {NoStop}%
\bibitem [{\citenamefont {Fedorenko}\ \emph {et~al.}(2014)\citenamefont
  {Fedorenko}, \citenamefont {Le~Doussal},\ and\ \citenamefont
  {Wiese}}]{fedorenko2014gaussian}%
  \BibitemOpen
  \bibfield  {author} {\bibinfo {author} {\bibfnamefont {A.~A.}\ \bibnamefont
  {Fedorenko}}, \bibinfo {author} {\bibfnamefont {P.}~\bibnamefont
  {Le~Doussal}}, \ and\ \bibinfo {author} {\bibfnamefont {K.~J.}\ \bibnamefont
  {Wiese}},\ }\href {\doibase 10.1209/0295-5075/105/16002} {\bibfield
  {journal} {\bibinfo  {journal} {EPL}\ }\textbf {\bibinfo {volume} {105}},\
  \bibinfo {pages} {16002} (\bibinfo {year} {2014})}\BibitemShut {NoStop}%
\bibitem [{\citenamefont {Fedorenko}\ and\ \citenamefont
  {Trimper}(2006)}]{fedorenko2006critical}%
  \BibitemOpen
  \bibfield  {author} {\bibinfo {author} {\bibfnamefont {A.~A.}\ \bibnamefont
  {Fedorenko}}\ and\ \bibinfo {author} {\bibfnamefont {S.}~\bibnamefont
  {Trimper}},\ }\href@noop {} {\bibfield  {journal} {\bibinfo  {journal}
  {Europhys. Lett.}\ }\textbf {\bibinfo {volume} {74}},\ \bibinfo {pages} {89}
  (\bibinfo {year} {2006})}\BibitemShut {NoStop}%
\bibitem [{\citenamefont {Itakura}(2003)}]{itakura2003frozen}%
  \BibitemOpen
  \bibfield  {author} {\bibinfo {author} {\bibfnamefont {M.}~\bibnamefont
  {Itakura}},\ }\href {\doibase 10.1103/PhysRevB.68.100405} {\bibfield
  {journal} {\bibinfo  {journal} {Phys. Rev. B}\ }\textbf {\bibinfo {volume}
  {68}},\ \bibinfo {pages} {100405} (\bibinfo {year} {2003})}\BibitemShut
  {NoStop}%
\bibitem [{\citenamefont {Feldman}(2002)}]{feldman2002critical}%
  \BibitemOpen
  \bibfield  {author} {\bibinfo {author} {\bibfnamefont {D.~E.}\ \bibnamefont
  {Feldman}},\ }\href {\doibase 10.1103/PhysRevLett.88.177202} {\bibfield
  {journal} {\bibinfo  {journal} {Phys. Rev. Lett.}\ }\textbf {\bibinfo
  {volume} {88}},\ \bibinfo {pages} {177202} (\bibinfo {year}
  {2002})}\BibitemShut {NoStop}%
\bibitem [{\citenamefont {Anfuso}\ and\ \citenamefont
  {Rosch}(2009)}]{anfuso2009random}%
  \BibitemOpen
  \bibfield  {author} {\bibinfo {author} {\bibfnamefont {F.}~\bibnamefont
  {Anfuso}}\ and\ \bibinfo {author} {\bibfnamefont {A.}~\bibnamefont {Rosch}},\
  }\href {\doibase 10.1140/epjb/e2009-00191-6} {\bibfield  {journal} {\bibinfo
  {journal} {Eur. Phys. J. B}\ }\textbf {\bibinfo {volume} {69}},\ \bibinfo
  {pages} {465} (\bibinfo {year} {2009})}\BibitemShut {NoStop}%
\bibitem [{\citenamefont {Nelson}\ and\ \citenamefont
  {Pelcovits}(1977)}]{nelson1977momentum}%
  \BibitemOpen
  \bibfield  {author} {\bibinfo {author} {\bibfnamefont {D.~R.}\ \bibnamefont
  {Nelson}}\ and\ \bibinfo {author} {\bibfnamefont {R.~A.}\ \bibnamefont
  {Pelcovits}},\ }\href {\doibase 10.1103/PhysRevB.16.2191} {\bibfield
  {journal} {\bibinfo  {journal} {Phys. Rev. B}\ }\textbf {\bibinfo {volume}
  {16}},\ \bibinfo {pages} {2191} (\bibinfo {year} {1977})}\BibitemShut
  {NoStop}%
\bibitem [{\citenamefont {Gamba}\ \emph {et~al.}(1999)\citenamefont {Gamba},
  \citenamefont {Grilli},\ and\ \citenamefont
  {Castellani}}]{gamba1999renormalization}%
  \BibitemOpen
  \bibfield  {author} {\bibinfo {author} {\bibfnamefont {A.}~\bibnamefont
  {Gamba}}, \bibinfo {author} {\bibfnamefont {M.}~\bibnamefont {Grilli}}, \
  and\ \bibinfo {author} {\bibfnamefont {C.}~\bibnamefont {Castellani}},\
  }\href {\doibase http://dx.doi.org/10.1016/S0550-3213(99)00340-5} {\bibfield
  {journal} {\bibinfo  {journal} {Nuclear Physics B}\ }\textbf {\bibinfo
  {volume} {556}},\ \bibinfo {pages} {463 } (\bibinfo {year}
  {1999})}\BibitemShut {NoStop}%
\end{thebibliography}

%

\end{document}